\documentclass[%
 reprint,
nofootinbib,
 amsmath,amssymb,
 aps,
]{revtex4-2}

\usepackage{graphicx}
\usepackage{dcolumn}
\usepackage{bm}
\usepackage[colorlinks=true,allcolors=blue]{hyperref}
\usepackage{makecell}
\usepackage{tabularx}
\usepackage{array}
\usepackage{nicefrac}
\newcolumntype{Y}{>{\centering\arraybackslash}X}
\renewcommand{\arraystretch}{1.2}
\usepackage{orcidlink}
\newcommand{\referee}[1]{\textcolor{black}{{#1}}}

\begin{document}

\preprint{APS/123-QED}

\title{The Simons Observatory: Combining cross-spectral foreground cleaning \\ with multitracer $B$-mode delensing for improved constraints on inflation}

\author{Emilie Hertig \orcidlink{0000-0001-9189-4035} $^{1,2}$}\email{emh83@cam.ac.uk}
\author{Kevin Wolz \orcidlink{0000-0003-3155-6151} $^{3}$}
\author{Toshiya Namikawa \orcidlink{0000-0003-3070-9240} $^4$}
\author{Antón Baleato Lizancos \orcidlink{0000-0002-0232-6480} $^{5,6,7}$}
\author{Susanna Azzoni \orcidlink{0000-0002-8132-4896} $^8$}
\author{Irene Abril-Cabezas \orcidlink{0000-0003-3230-4589} $^{9,2}$}
\author{David Alonso \orcidlink{0000-0002-4598-9719} $^3$}
\author{Carlo Baccigalupi \orcidlink{0000-0002-8211-1630} $^{10,11,12,13}$}
\author{Erminia Calabrese \orcidlink{0000-0003-0837-0068} $^{14}$}
\author{Anthony Challinor \orcidlink{0000-0003-3479-7823} $^{1,2,9}$}
\author{Josquin Errard \orcidlink{0000-0002-1419-0031} $^{15}$}
\author{Giulio Fabbian \orcidlink{0000-0002-3255-4695} $^{14}$}
\author{Carlos Herv\'ias-Caimapo \orcidlink{0000-0002-4765-3426} $^{16}$}
\author{Baptiste Jost \orcidlink{0000-0002-0819-751X} $^{17,4}$}
\author{Nicoletta Krachmalnicoff $^{10,11,13}$}
\author{Anto I. Lonappan \orcidlink{0000-0003-1200-9179} $^{18}$}
\author{Magdy Morshed $^{15,19}$}
\author{Luca Pagano \orcidlink{0000-0003-1820-5998} $^{20,21,22}$}
\author{Blake Sherwin \orcidlink{0000-0002-4495-1356} $^{9,2}$}

\affiliation{$^1$Institute of Astronomy, University of Cambridge, Madingley Road, Cambridge, CB3 0HA, UK}
\affiliation{$^2$Kavli Institute for Cosmology Cambridge, Madingley Road, Cambridge, CB3 0HA, UK}
\affiliation{$^3$Department of Physics, University of Oxford, Denys Wilkinson Building, Keble Road, Oxford, OX1 3RH, UK}
\affiliation{$^4$Center for Data-Driven Discovery, Kavli IPMU (WPI), UTIAS, The University of Tokyo, Kashiwa, 277-8583, Japan}
\affiliation{$^5$Berkeley Center for Cosmological Physics, UC Berkeley, CA 94720, USA}
\affiliation{$^6$Department of Physics, UC Berkeley, CA 94720, USA}
\affiliation{$^7$Lawrence Berkeley National Laboratory, One Cyclotron Road, Berkeley, CA 94720, USA}
\affiliation{$^8$Department of Astrophysical Sciences, Peyton Hall, Princeton University, Princeton, NJ 08544, USA}
\affiliation{$^9$DAMTP, University of Cambridge, Cambridge, CB3 0WA, UK}
\affiliation{$^{10}$The International School for Advanced Studies (SISSA), via Bonomea 265, I-34136 Trieste, Italy}
\affiliation{$^{11}$The National Institute for Nuclear Physics (INFN), via Valerio 2, I-34127, Trieste, Italy}
\affiliation{$^{12}$The National Institute for Astrophysics (INAF), via Tiepolo 11, I-34143, Trieste, Italy}
\affiliation{$^{13}$The Institute for Fundamental Physics of the Universe (IFPU), Via Beirut 2, I-34151, Trieste, Italy}
\affiliation{$^{14}$School of Physics and Astronomy, Cardiff University, The Parade, Cardiff, Wales CF24 3AA, UK}
\affiliation{$^{15}$Universit\'e Paris Cit\'e, CNRS, Astroparticule et Cosmologie, F-75013 Paris, France}
\affiliation{$^{16}$ Instituto de Astrof\'isica and Centro de Astro-Ingenier\'ia, Facultad de F\'isica, \\ Pontificia Universidad Cat\'olica de Chile, Av. Vicu\~na Mackenna 4860, 7820436 Macul, Santiago, Chile}
\affiliation{$^{17}$Kavli Institute for the Physics and Mathematics of the Universe (Kavli IPMU, WPI), UTIAS, The University of Tokyo, Kashiwa, Chiba 277-8583, Japan}
\affiliation{$^{18}$Dipartimento di Fisica, Universita di Roma Tor Vergata, Via della Ricerca Scientifica, 1, 00133 Roma, Italy}
\affiliation{$^{19}$CNRS-UCB International Research Laboratory Centre Pierre Binétruy, CPB-IN2P3, Berkeley, US}
\affiliation{$^{20}$Dipartimento di Fisica e Scienze della Terra, Universit\`a degli Studi di Ferrara, via Saragat 1, I-44122 Ferrara, Italy}
\affiliation{$^{21}$Istituto Nazionale di Fisica Nucleare, Sezione di Ferrara, via Saragat 1, I-44122 Ferrara, Italy}
\affiliation{$^{22}$Institut d'Astrophysique Spatiale, CNRS, Univ. Paris-Sud, Univ. Paris-Saclay, 91405 Orsay cedex, France}

\collaboration{The Simons Observatory Collaboration}

\date{\today}

\begin{abstract} 
    The Simons Observatory (SO), due to start full science operations in early 2025, aims to set tight constraints on inflationary physics by inferring the tensor-to-scalar ratio $r$ from measurements of CMB polarization $B$-modes. Its nominal
    design including three small-aperture telescopes (SATs) targets a precision $\sigma(r=0)\leq0.003$ without delensing. Achieving this goal and further reducing uncertainties requires a thorough understanding and mitigation of other large-scale $B$-mode sources such as Galactic foregrounds and weak gravitational lensing. We present an analysis pipeline aiming to estimate $r$ by including delensing within a cross-spectral likelihood, and demonstrate it for the first time on SO-like simulations accounting for various levels of foreground complexity, inhomogeneous noise and partial sky coverage. As introduced in an earlier SO delensing paper, lensing $B$-modes are synthesised using internal CMB lensing reconstructions as well as Planck-like cosmic infrared background maps and LSST-like galaxy density maps. 
    We then extend SO’s power-spectrum-based foreground-cleaning algorithm to include all auto- and cross-spectra between the lensing template and the SAT $B$-modes in the likelihood function. This allows us to constrain $r$ and the parameters of our foreground model simultaneously.
    Within this framework, we demonstrate the equivalence of map-based and cross-spectral delensing and use it to motivate an optimized pixel-weighting scheme for power spectrum estimation. We start by validating our pipeline in the simplistic case of uniform foreground spectral energy distributions (SEDs). In the absence of primordial $B$-modes, we find that the $1\sigma$ statistical uncertainty on $r$, $\sigma(r)$, decreases by 37\% as a result of delensing. Tensor modes at the level of $r=0.01$ are successfully detected by our pipeline. Even when using more realistic foreground models including spatial variations in the dust and synchrotron spectral properties, we obtain unbiased estimates of $r$ both with and without delensing by employing the moment-expansion method. In this case, uncertainties are increased due to the higher number of model parameters, and delensing-related improvements range between 27\% and 31\%. These results constitute the first realistic assessment of the delensing performance at SO’s nominal sensitivity level.
\end{abstract}

\maketitle


\section{Introduction}

Anisotropies of the cosmic microwave background (CMB) offer a unique window on the first instants after the birth of the Universe, probing fundamental physics at energy scales unreachable to any earthbound accelerator. After Planck's cosmic-variance-limited measurements of the temperature power spectrum \cite{planck_collaboration_planck_2020}, present and future CMB experiments are directing their focus towards polarization patterns; large-scale $B$-modes are particularly interesting as they hold the key to a potential first detection of primordial gravitational waves (PGW). 
Predicted by most inflation models, these tensor fluctuations of the spacetime metric are the only ones capable of producing parity-odd polarization anisotropies, unlike the scalar (density) perturbations that exclusively create $E$-modes at linear order \cite{kamionkowski_statistics_1997,seljak_signature_1997}. The detection of primordial $B$-modes would therefore constitute a significant breakthrough, providing unprecedented evidence in favour of the inflationary scenario, which, despite being widely accepted as a theoretical framework due to its ability to generate the right initial conditions for our Universe, still lacks direct experimental verification.

Two recent studies combining maps from BICEP/Keck, Planck and WMAP with baryon acoustic oscillation (BAO) data allowed to set the tightest constraints to date on the amplitude of PGW, as described by the tensor-to-scalar ratio $r$: they respectively inferred $r<0.036$ \cite{bicepkeck_collaboration_bicep_2021} and $r<0.032$ \cite{tristram_improved_2022} at 95\% confidence for a pivot scale of $0.05\,\text{Mpc}^{-1}$. Together with the latest bounds on the scalar spectral index of the power spectrum of primordial curvature perturbations,
these results favour a concave potential for the inflaton field driving inflation and have ruled out monomial models as well as natural inflation \cite{bicepkeck_collaboration_bicep_2021}. While many inflationary scenarios remain viable \cite{martin_encyclopaedia_2023}, the absence of a detection of PGW at the sensitivity level of current experiments has also motivated the development of alternative theories aiming to explain observed features of the Universe without the need for such a mechanism. These include, for example, bouncing cosmologies \cite{ijjas_bouncing_2018} as well as CPT-symmetric models \cite{boyle_cpt-symmetric_2018}. New measurements of $r$ with lower uncertainties will be an essential step towards discriminating between this wide range of possible descriptions of the early Universe.

The Simons Observatory (SO) \cite{the_simons_observatory_collaboration_simons_2019}, whose construction on Cerro Toco in the Atacama Desert is nearing completion with full science operations expected to start in early 2025, will significantly improve upon existing constraints: with deep observations of around $10\%$ of the sky by three $0.5\,\text{m}$ small-aperture telescopes (SATs), its nominal design aims to detect or rule out $r \geq 0.01$ at the $3\sigma$ level \cite{the_simons_observatory_collaboration_simons_2019-1}. The planned addition of three more SATs 
will allow to reduce error bars even further. SO will therefore set tight bounds on currently allowed inflation scenarios such as quartic hilltop potentials and $\alpha$-attractors~\cite{linde_single-field_2015}. 
Furthermore, the technologies and analysis pipelines developed for SO pave the way towards a new generation of experiments including CMB-S4 \cite{abazajian_cmb-s4_2022} and LiteBIRD \cite{litebird_collaboration_probing_2023}, which will push limits to $\sigma(r) < 10^{-3}$ in the early 2030s. This foreshadows a particularly exciting time for early-Universe physics as several popular models, for example Starobinsky \cite{starobinskii_spectrum_1979} and Higgs inflation \cite{bezrukov_standard_2008}, will then be decisively tested. 

In order to reach its target precision, SO will have to overcome two main challenges. The first relates to the weak gravitational lensing of CMB photons by large-scale structures \cite{lewis_weak_2006}, which sources large-scale $B$-modes from intermediate- and small-scale $E$-modes~\cite{hanson_detection_2013}. Consequently, the deflected CMB light reaching our detectors contains an additional, lensing $B$-mode component sourced by scalar fluctuations rather than PGW, whose amplitude surpasses that of the primordial tensor signal. This lensing-induced noise
is larger than instrumental noise at SO's sensitivity level, and its contribution to sample variance will therefore be one of the main limiting factors affecting parameter constraints. Mitigating this effect, a process known as delensing, requires an accurate model of the specific lensing $B$-mode realization observed in our sky. Such a template is built by convolving measurements of intermediate- and small-scale $E$-modes with an estimator of the CMB lensing potential. In SO's case, both of these products will be extracted from high-resolution maps obtained with a $6\,\text{m}$ large-aperture telescope (LAT); external large-scale structure tracers such as the cosmic infrared background (CIB) and galaxy surveys will also contribute to reconstructing the lensing convergence \cite{namikawa_simons_2022}. Delensing is performed by subtracting the $B$-mode template from the observed polarization maps, or by cross-correlating it with the data. This parametric power-spectrum-based approach was recently demonstrated for the first time by the BICEP/Keck Collaboration and resulted in a 10\% improvement on the inferred $r$ constraint \cite{bicepkeck_and_sptpol_collaborations_demonstration_2021}. While the present work uses the same component separation technique, delensing of SO data is predicted to lead to a greater decrease in $\sigma(r)$ due to lower noise levels.

The second main challenge encountered by PGW searches is due to polarized Galactic foregrounds, in particular thermal dust emission and synchrotron radiation \cite{choi_polarized_2015}. For the $r$ values targeted by SO, the total $B$-modes produced by these sources dominate over the predicted inflationary tensor polarization pattern by orders of magnitude; on the angular scales of interest, their combined amplitude is at least equivalent to that of a primordial signal with $r\sim0.05$ for any part of the sky~\cite{krachmalnicoff_characterization_2016}. Distinguishing these contaminants from the CMB is therefore an essential step towards inferring accurate cosmological information from the data \cite{delabrouille_diffuse_2007}. Existing foreground cleaning techniques rely on the distinctive spectral energy distributions (SEDs) of Galactic emissions, which differ from the CMB blackbody spectrum and can be separated from the latter by multifrequency observations \cite{leach_component_2008}. Three independent algorithms have been designed to perform this task using SO's six frequency bands (from 27 to $280\,\text{GHz}$), both at the map and power spectrum levels, and have been tested on realistic SO-like simulations recently~\cite{wolz_simons_2023}. Delensing was not performed at this stage; its effect was instead estimated by using input simulations with decreased lensing $B$-mode power. 

The present work aims to include delensing in the power-spectrum-based parametric component separation pipeline from Ref.~\cite{wolz_simons_2023}, optimize its performance and characterize the subsequent reduction of statistical uncertainties on $r$ when applied to simulations mimicking SO's noise and foreground properties. Lensing $B$-mode templates are obtained using the multitracer approach described in Ref.~\cite{namikawa_simons_2022}, where cross-spectral delensing was demonstrated for an idealistic foreground-free case. Different levels of foreground complexity are investigated; biases related to residuals from spatial variability of foreground SEDs are mitigated by using a technique known as `moment expansion', which introduces additional parameters to model deviations from the sky average of the spectral properties~\cite{tegmark_removing_1998,chluba_rethinking_2017,vacher_high_2023}. This method was first demonstrated for SO $B$-modes in Ref.~\cite{azzoni_minimal_2021}.

The paper is structured as follows. Section~\ref{section:theory} summarizes the working principle of our analysis pipeline as well as the theoretical and mathematical foundations on which it relies. In Sec.~\ref{section:equivalence}, we demonstrate the equivalence between map-based methods and the cross-spectral approach adopted here. Building upon this result, we optimize the delensing performance in Sec.~\ref{section:weighting} by defining a new pixel-weighting scheme accounting for both instrumental noise and lensing $B$-mode variance. Section~\ref{section:pipeline} describes the input simulations and the practical implementation of our pipeline, while statistical uncertainties and biases on $r$ for all considered foreground and instrumental noise models are presented and analyzed in Sec.~\ref{section:results}. Finally, we conclude and explore prospects for future work in Sec.~\ref{section:conclusion}. Further details on the likelihood derivation and tests performed with an alternative pixel-weighting scheme are provided in Appendices~\ref{appendix_A} and~\ref{appendix_B} respectively.

\section{Theoretical framework}\label{section:theory}

\subsection{Delensing: motivation and principle}

We start by outlining essential mathematical results related to CMB lensing and polarization, in order to illustrate the importance of delensing and characterize the $B$-mode template construction stage of our pipeline.

\subsubsection{CMB lensing and cosmic variance}

The linear polarization of light is quantified by the Stokes parameters $Q$ and $U$, which are the components of a symmetric trace-free tensor on the sphere $\mathcal{P}_{ab}$ such that $\mathcal{P}_{\hat{\theta}\hat{\theta}}=-\mathcal{P}_{\hat{\phi}\hat{\phi}}=Q/2$ and $\mathcal{P}_{\hat{\theta}\hat{\phi}}=\mathcal{P}_{\hat{\phi}\hat{\theta}}=U/2$ in a right-handed orthonormal basis\footnote{Here, $\hat{\bm{n}}$ is the line-of-sight direction and $\hat{\bm{\theta}}$ and $\hat{\bm{\phi}}$ are along the $\theta$ and $\phi$ directions of a spherical coordinate system.}
$(\hat{\bm{n}},\hat{\bm{e}}_{\theta},\hat{\bm{e}}_{\phi})$. As photons travelling from the last-scattering surface are deflected by potential wells along the line of sight $\hat{\bm{n}}$, $\mathcal{P}_{ab}$ is subjected to the direction-dependent remapping 
\begin{equation}\label{eq:remapping}
    \tilde{\mathcal{P}}_{ab}(\hat{\bm{n}})=\mathcal{P}_{ab}(\hat{\bm{n}})+\bm{\alpha}^c(\hat{\bm{n}})\nabla_c\mathcal{P}_{ab}(\hat{\bm{n}})
\end{equation}
at leading order, where the tilde refers to lensed quantities \cite{challinor_geometry_2002}. In the Born approximation, the small deflection angle $\bm{\alpha}(\hat{\bm{n}})$ corresponds to the gradient of the lensing potential $\phi$. Using the spin-weight formalism described in \cite{lewis_analysis_2001}, the Stokes parameters are related to $E$- and $B$-modes as follows: 
\begin{equation}\label{eq:harmonic_int}
    E_{lm}\pm iB_{lm}=-\int{(Q\pm iU)(\hat{\bm{n}})_{\pm2}Y^{*}_{lm}(\hat{\bm{n}})d^2\hat{\bm{n}}}.
\end{equation}
Integrating Eq.~\eqref{eq:remapping} allows one to derive the lensing $B$-mode contribution in the absence of primordial tensor perturbations \cite{namikawa_simons_2022}:
\begin{equation}\label{eq:lensing_B}
    B^{\textrm{lens}}_{lm}=-i\sum_{l',m'}{\sum_{L,M}{\left(\begin{matrix} l & l'& L \\ m & m' & M \end{matrix}\right)p^{-}F^{(2)}_{ll'L}E^{*}_{l'm'}\kappa^{*}_{LM}}}.
\end{equation}
In Eq.~\eqref{eq:lensing_B}, the factor $p^{-}$ is 0 for $l+l'+L$ even and 1 for $l+l'+L$ odd, while $\kappa=-\frac{1}{2}\nabla^2\phi$ represents the lensing convergence and the spin-$s$ mode coupling function $F^{(s)}_{ll'L}$ is given by
\begin{multline}\label{eq:mode_coupling}
    F^{(s)}_{ll'L}=\frac{2}{L(L+1)}\left[l'(l'+1)+L(L+1)-l(l+1)\right]
    \\
    \times \sqrt{\frac{(2l+1)(2l'+1)(2L+1)}{16\pi}}\left(\begin{matrix} l & l'& L \\ s & -s & 0 \end{matrix} \right).
\end{multline}
As a result of this coupling, large-scale lensing $B$-modes receive contributions from $E$-modes at all scales, in particular from the high multipoles at which their power spectrum peaks \cite{lewis_weak_2006}.

The presence of an additional $B$-mode component in the data increases cosmic variance, thus amplifying uncertainties on inferred parameters. As an illustration, let us assume that the observed $B$-modes, $B_{lm}$, are a full-sky Gaussian field such that $\langle B^{*}_{lm}B_{l'm'}\rangle=\delta_{ll'}\delta_{mm'}C_{l}$, where the angle brackets denote an ensemble average over CMB and noise realizations. We fit this power spectrum with a theoretical model $rC_{l}^{\textrm{prim}}+C_{l}^{\textrm{lens}}+N_{l}$ consisting of the primordial signal, the lensing term\footnote{Since the lensing $B$-mode power spectrum is predicted very accurately within a given cosmological model and the latter is much better constrained
than the foregrounds, we can afford to treat it as a fixed
component. As discussed in Sec.~\ref{section:conclusion}, the possible inclusion of a free parameter modulating the lensing amplitude when working with real data is not expected to significantly impact $\sigma(r)$.} and a noise component. The full-sky log likelihood at a given multipole $l$ corresponds to 
\begin{align}
-\ln{\mathcal{L}(r)} &= \frac{1}{2}\sum_m{\left(B^{*}_{lm}B_{lm}C_{l}^{-1} + \ln{C_{l}}\right)} \nonumber \\ &= \frac{2l+1}{2} \left(\hat{C}_l C_l^{-1} + \ln{C_l}\right),
\label{eq:gaussian}
\end{align}
where the power spectrum estimator satisfies $\langle\hat{C}_l\rangle=C_l$. For $r=0$, the associated statistical error can then be derived from the Fisher information:
\begin{align}
\left. \sigma^{-2}(r)\right|_{r=0}&=\sum_{l}
\left. \left\langle-\frac{\partial^2 \ln{\mathcal{L}}}{\partial r^2}\right\rangle\right|_{r=0}\nonumber \\
&=\sum_{l}\left(\frac{2l+1}{2}\right)\left(\frac{C_l^{\textrm{prim}}}{C_l^{\textrm{lens}}+N_l}\right)^2 .
\label{eq:sigma_r_fisher}
\end{align}

The lensing $B$-mode contribution to Eq.~\eqref{eq:sigma_r_fisher}, which behaves similarly to white noise on large scales with standard deviation of approximately
$5\,\mu\text{K-arcmin}$, already surpasses SO's goal noise level of around $2\,\mu\text{K-arcmin}$ and therefore constitutes a significant limitation. While the heuristic argument above does not capture complex effects such as survey non-idealities (e.g., masking, inhomogeneous noise and foreground residuals) or the non-Gaussianity of lensing $B$-modes, it highlights the importance of delensing for SO and future PGW experiments.

\subsubsection{$B$-mode template construction}
In order to reduce the lensing contribution to the variance of $r$ measurements, a template matching the particular realization of the lensing $B$-modes on our sky must be computed and subtracted from the data. This is done by evaluating Eq.~\eqref{eq:lensing_B} for estimates of the $E$-modes and the lensing convergence; Ref.~\cite{namikawa_simons_2022} previously investigated the lensing $B$-mode template construction process for SO. In the following, we summarize the method used to extract the $E$-modes and the lensing potential and comment on the differences between the present paper and Ref.~\cite{namikawa_simons_2022}. Note that the multifrequency LAT data used at this stage will be subjected to map-based foreground cleaning before any information is drawn from it; in the present analysis, we assume foreground residuals to be negligible and do not include them in our simulations. Investigating the impact of such residuals will be the object of future work (see Sec.~\ref{section:conclusion}).

\begin{figure*}[!t]
    \centering
    \begin{minipage}{.48\textwidth}
        \centering
        \includegraphics[width=\linewidth, height=0.23\textheight]{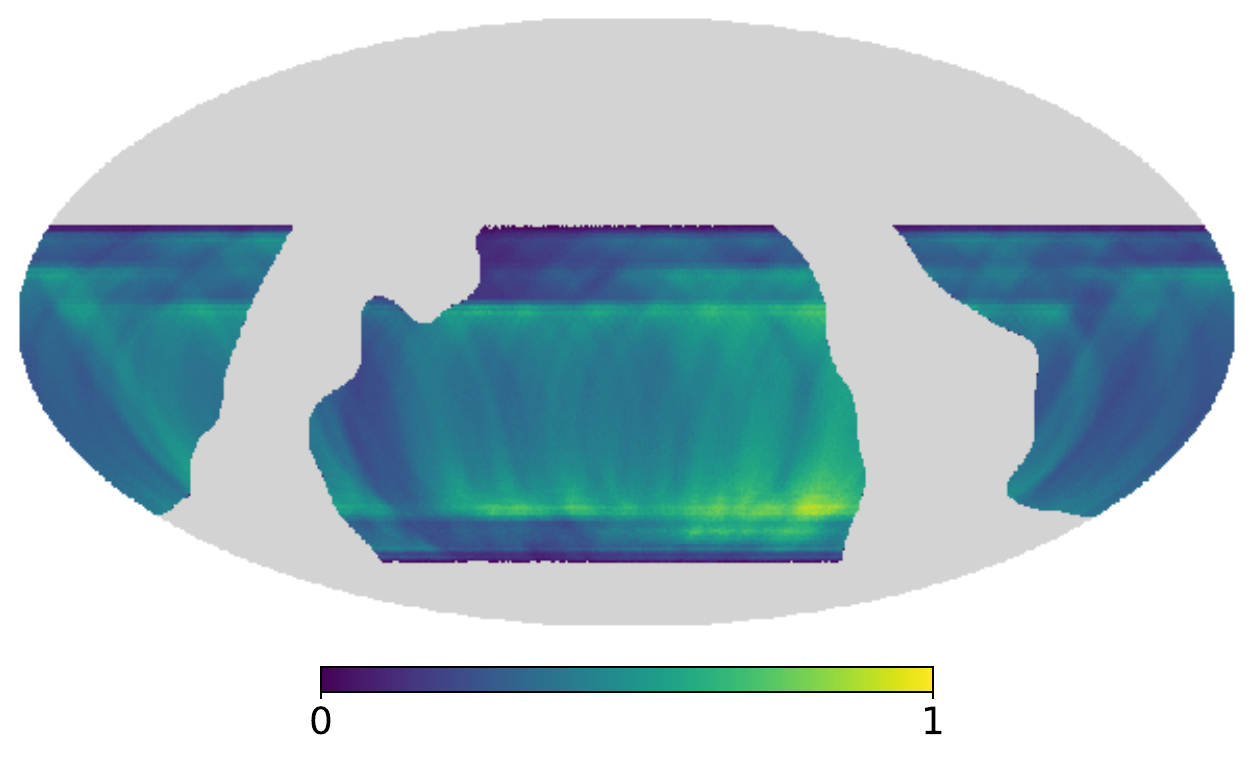}
    \end{minipage}%
    \hfill
    \begin{minipage}{0.48\textwidth}
        \centering
        \includegraphics[width=\linewidth, height=0.23\textheight]{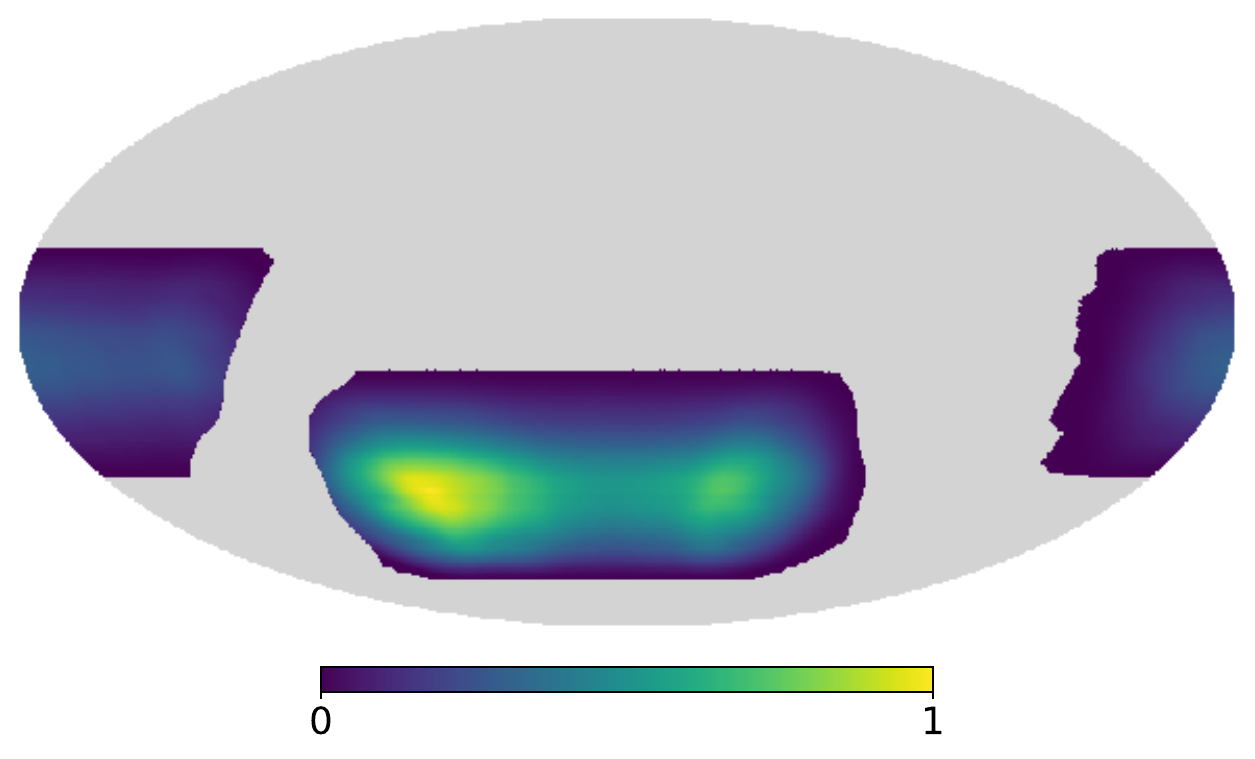}
    \end{minipage}
    \vspace{-5pt}
    \caption{Normalized hit counts over the areas observed by the LAT (left) and SATs (right) in equatorial coordinates, multiplied by the Planck 70\% Galactic mask. Masked pixels are shown in grey. These maps were made using the latest version of the planned SO scanning strategy. Of the pixels observed by the SAT, 96\% are also seen by the LAT as a result of using slightly narrower Galactic cuts than in Ref.~\cite{wolz_simons_2023} to maximize overlap.}
    \label{fig:hitmaps}
\end{figure*}

Small-scale $E$-modes are obtained by coadding data from three LAT frequency channels (93, 145 and 225\,GHz) with inverse-noise-variance weighting in harmonic space. Instrumental noise is then mitigated by applying a diagonal Wiener filter
\begin{equation}\label{eq:wiener}
\hat{E}_{lm}^{\rm{WF}} = \frac{C_l^{EE}}{C_l^{EE}+N_l^{EE}} \hat{E}_{lm},
\end{equation}
where $C_l^{EE}$ is the theoretical lensed $E$-mode power spectrum, $N_l^{EE}$ is computed for the coadded bands using the LAT noise model and hats refer to observed quantities. In Ref.~\cite{namikawa_simons_2022}, the Wiener-filtered $E$-modes are obtained with the algorithm presented in Ref.~\cite{eriksen_power_2004} to account for the inhomogeneities of noise and the survey boundary. Furthermore, Ref.~\cite{namikawa_simons_2022} includes SAT $E$-modes, defined over the smaller area displayed in the right panel of Fig.~\ref{fig:hitmaps}. This technique was shown to improve the correlation between the estimator and the input $E$-modes for the foreground-free situation studied in Ref.~\cite{namikawa_simons_2022}, but has the disadvantage of being very CPU intensive. To reduce computational costs, the present analysis will be restricted to LAT-only diagonally filtered $E$-modes. This condition also prevents our lensing templates from containing SAT foreground residuals and instrument systematics; these would contribute as small corrections to the cross-spectra between the template and the SAT maps, and will be the subject of future work. Note that while Eq.~\eqref{eq:lensing_B} refers to unlensed $E$-modes, the latter are not directly observable and our estimator is built using their lensed counterparts; gradient-order templates computed this way have actually been found to result in more efficient delensing due to cancellations between higher-order terms in the $B$-mode residuals \cite{lizancos_limitations_2021}.

Let us now briefly describe the method used to estimate the lensing convergence. As was done in Ref.~\cite{namikawa_simons_2022}, we combine the LAT CMB lensing map with external large-scale structure tracers. To reconstruct the CMB lensing map internally, we use quadratic estimators based on the correlation between lensed temperature ($\Theta$) and polarization ($E$, $B$) fields, whose ensemble average over \referee{unlensed CMB fields but with a fixed realisation of the $\kappa$ field} is given by
\begin{multline}\label{eq:off_diag}
    \langle \tilde{X}_{lm} \tilde{Y}_{l’m’} \rangle = \delta_{ll'}\delta_{m-m'}(-1)^{m}C_{l}^{XY} \\ + \sum_{L,M}{\left(\begin{matrix} l & l'& L \\ m & m' & M \end{matrix}\right)f^{XY}_{ll'L}\kappa^{*}_{LM}}
\end{multline}
at leading order \cite{namikawa_lensing_2014}. The second (off-diagonal) term in Eq.~\eqref{eq:off_diag} appears as a result of mode coupling and the response functions $f^{XY}_{ll'L}$ correspond to 
\begin{equation}\label{eq:response_functions}
    \begin{split}
        f^{\Theta \Theta}_{ll'L} &= F^{(0)}_{ll'L}C_{l'}^{\Theta \Theta} + F^{(0)}_{l'lL}C_{l}^{\Theta \Theta} \\
        f^{\Theta E}_{ll'L} &= p^{+}F^{(0)}_{ll'L}C_{l'}^{\Theta E} + p^{+}F^{(2)}_{l'lL}C_{l}^{\Theta E} \\
        f^{E E}_{ll'L} &= p^{+}F^{(2)}_{ll'L}C_{l'}^{E E} + p^{+}F^{(2)}_{l'lL}C_{l}^{E E} \\
        f^{E B}_{ll'L} &= -i p^{-}F^{(2)}_{l'lL}C_{l}^{E E},
    \end{split}
\end{equation}
where $p^+$ is $1$ for $l+l'+L$ even and $0$ for $l+l'+L$ odd.
Aiming to build a robust estimator of $\kappa$ satisfying $\langle\hat{\kappa}_{LM}\rangle=\kappa_{LM}$, Ref.~\cite{okamoto_cmb_2003} proposed the form 
\begin{equation}\label{eq:kappa_estimator}
    \left(\hat{\kappa}^{XY}_{LM}\right)^{*} = A_L^{XY}\sum_{ll'mm'}\left(\begin{matrix} l & l'& L \\ m & m' & M \end{matrix}\right)\frac{(f^{XY}_{ll'L})^{*}}{\Delta^{XY}}\bar{X}_{lm}\bar{Y}_{l'm'},
\end{equation}
where $\Delta^{XY}\big|_{X\neq Y}=1$, $\Delta^{XY}\big|_{X=Y}=2$ and the fields $\bar{X}_{lm}$ and $\bar{Y}_{l'm'}$ are the inverse-variance-filtered observed fields, related to the Wiener-filtered fields in Eq.~\eqref{eq:wiener} by $\bar{X}_{lm} = \hat{X}_{lm}^{\text{WF}}/C_l^{XX}$.
Taking the ensemble average of Eq.~\eqref{eq:kappa_estimator} and substituting Eq.~\eqref{eq:off_diag} for $\langle \hat{X}_{lm} \hat{Y}_{l’m’} \rangle$, we find that the contribution from the diagonal term vanishes for $L>0$ and the off-diagonal term allows to determine the normalization factor
\begin{equation}\label{eq:normalization}
    A_L^{XY} = \left[\sum_{ll'}{\frac{(2L+1)^{-1}|f^{XY}_{ll'L}|^2}{\Delta^{XY}\left(C_l^{XX}+N_l^{XX}\right)\left(C_{l'}^{YY}+N_{l'}^{YY}\right)}}\right]^{-1}.
\end{equation}
Again, Eq.~\eqref{eq:response_functions} refers to unlensed fields, however the lensed ones used in practice have been found to provide a better approximation to the non-perturbative result for correlators involving $\hat{\kappa}_{LM}$ \cite{lewis_shape_2011}.

The expression \eqref{eq:kappa_estimator} with $A_L^{XY}$ given by Eq.~\eqref{eq:normalization} corresponds to the quadratic estimator (QE) of the lensing convergence as described in Ref.~\cite{okamoto_cmb_2003}. All field pairings listed in Eq.~\eqref{eq:response_functions} ($\Theta\Theta$, $\Theta E$, $EE$ and $EB$) are included in the present work, and are coadded with weights minimizing the reconstruction noise variance at each multipole.
We do not use the \referee{$\Theta B$} combination due to its low signal-to-noise ratio. As explained in Refs~\cite{namikawa_cmb_2017} and \cite{lizancos_impact_2021}, the $EB$ quadratic estimator becomes a source of bias if the multipole range of the fields used to build it overlaps with that of the $B$-modes we aim to delens. We therefore only consider modes between $l=301$ and $l=4096$ in our lensing reconstruction; the temperature field is further restricted to $500 < l < 3000$ in order to avoid large-scale atmospheric noise and residual small-scale extragalactic foreground contributions \cite{van_engelen_cmb_2014,lizancos_impact_2022}. The latter are not included in our simulations at this stage, but will be important for the analysis of real SO data. Finally, note that noise inhomogeneities and the presence of a survey mask can cause the ensemble average $\langle\hat{\kappa}^{XY}_{LM}\rangle$ to become non-zero. Following Ref.~\cite{namikawa_simons_2022}, we mitigate this bias for each quadratic estimator by subtracting a mean-field correction computed from the average of our simulations.

Combining all possible internal tracers of $\kappa$ still does not lead to optimal delensing performance. Indeed, reconstruction noise resulting from statistical fluctuations of the unlensed CMB is a significant limitation at intermediate and high multipoles which are particularly relevant for delensing. Quadratic estimators should therefore be complemented by external large-scale structure tracers such as galaxy surveys or measurements of the CIB. For future SO analysis, the latter will be extracted from Planck data using, for example, the GNILC algorithm~\cite{planck_collaboration_planck_2016_1}, while the DESI Legacy Imaging Surveys~\cite{dey_overview_2019}, unWISE~\cite{schlafly_unwise_2019} and the upcoming LSST~\cite{ivezic_lsst_2018} will provide information on the spatial distribution of galaxies.
Other types of internal lensing reconstruction techniques, for example likelihood-based estimators~\cite{hirata_reconstruction_2003}, are not included in this work as they are not expected to outperform QEs significantly for SO's nominal configuration~\cite{the_simons_observatory_collaboration_simons_2019}.

Coadding all tracers $\hat{\kappa}^{i}$ with weights chosen to maximize the correlation between our final estimator $\hat{\kappa}^{\textrm{comb}}$ and the true lensing convergence $\kappa$, we obtain\footnote{Here, we ignore correlations between multipoles. Further optimization would be possible by accounting for such effects as was done in Ref.~\cite{namikawa_litebird_2023}. However, in our setup, the external tracers are Gaussian fields with no correlations between multipoles, so the delensing efficiency from simulations is close to that obtained with idealized forecasts; extending Eq.~\eqref{eq:kappa_comb} would thus not significantly improve $\sigma(r)$.} \cite{sherwin_delensing_2015}
\begin{equation}\label{eq:kappa_comb}
    \hat{\kappa}^{\textrm{comb}}_{LM}=\sum_{ij}(\rho^{-1})^{ij}_{L}\rho^{j\kappa}_{L}\sqrt{\frac{C_L^{\kappa\kappa}}{C_L^{\hat{\kappa}^{i}\hat{\kappa}^{i}}}}\hat{\kappa}^{i}_{LM}.
\end{equation}
In Eq.~\eqref{eq:kappa_comb}, $\rho^{ij}_L$ and $\rho^{i\kappa}_L$ represent, respectively, correlation coefficients between tracers $\hat{\kappa}^{i}$ and $\hat{\kappa}^{j}$, or between $\hat{\kappa}^{i}$ and the true lensing convergence. These quantities, which are the components of the correlation matrix $\rho_L$, were computed in Ref.~\cite{yu_multitracer_2017} for the simulated tracers used in this work. As shown in Fig.~2 of Ref.~\cite{namikawa_simons_2022}, internal reconstruction is the most relevant tracer for $L < 250$; at smaller scales ($250 < L < 1000$), both the CIB and galaxy overdensity maps have a higher degree of correlation with the true convergence for the noise level of the SO LAT.

The lensing $B$-mode template, built by performing the weighted convolution of $\hat{E}_{lm}^{\rm{WF}}$ (with $50 < l < 2048$) and $\hat{\kappa}_{LM}^{\textrm{comb}}$ (with $20 < L < 2048$) as indicated in Eq.~\eqref{eq:lensing_B}, can then be subtracted from observations at the map level (over the regions observed by both the SATs and the LAT) or cross-correlated with the data in order to infer tighter constraints on $r$. While this paper focuses on the latter approach, both techniques will ultimately be used for SO and are expected to yield equivalent results (see Sec.~\ref{section:equivalence}).

Finally, note that the ensemble-average cross-spectra of the lensing template with SAT $B$-modes are identical to the template auto-spectrum for ideal (statistically isotropic) surveys. Indeed, Eq.~\eqref{eq:kappa_comb} implies 
\begin{equation}\label{eq:kappa_comb_spectra}
C_L^{\kappa\hat{\kappa}^{\textrm{comb}}} = C_L^{\hat{\kappa}^{\textrm{comb}}\hat{\kappa}^{\textrm{comb}}} = C_L^{\kappa\kappa}\sum_{ij}\rho^{i\kappa}_{L}(\rho^{-1})^{ij}_{L}\rho^{j\kappa}_{L}.
\end{equation}
Considering the Wiener filter in Eq.~\eqref{eq:wiener} and the fact that $C_l^{E\hat{E}}=C_l^{EE}$ as the noise component is uncorrelated with the underlying $E$-modes, we then obtain the following expression for the cross-spectrum between the template (denoted by the superscript $B_t$) and the true lensing $B$-modes:
\begin{align}
    C^{B_{t}B}_l &= \sum_{l'L}{|\mathcal{M}(l,l',L)|^2 \frac{C^{EE}_{l'}}{C^{\hat{E}\hat{E}}_{l'}}C^{E\hat{E}}_{l'}
    C^{\kappa\hat{\kappa}^{\textrm{comb}}}_{L}} \nonumber \\
    &= \sum_{l'L}{|\mathcal{M}(l,l',L)|^2 \frac{C^{EE}_{l'}}{C^{\hat{E}\hat{E}}_{l'}}C^{EE}_{l'}C^{\kappa\hat{\kappa}^{\textrm{comb}}}_{L}}, \label{eq:Cl_Bxtemp}
\end{align}
where all the prefactors from Eq.~\eqref{eq:lensing_B} have been grouped into $\mathcal{M}(l,l',L)$ for legibility. Similarly, the lensing template auto-spectrum is given by
\begin{align}
    C^{B_{t}B_{t}}_l &= \sum_{l'L}{|\mathcal{M}(l,l',L)|^2 \left(\frac{C^{EE}_{l'}}{C^{\hat{E}\hat{E}}_{l'}}\right)^2C^{\hat{E}\hat{E}}_{l'}C^{\hat{\kappa}^{\textrm{comb}}\hat{\kappa}^{\textrm{comb}}}_{L}} \nonumber \\
    &= \sum_{l'L}{|\mathcal{M}(l,l',L)|^2 \frac{C^{EE}_{l'}}{C^{\hat{E}\hat{E}}_{l'}}C^{EE}_{l'}C^{\kappa\hat{\kappa}^{\textrm{comb}}}_{L}}. \label{eq:Cl_tempxtemp}
\end{align}

The equality between Eqs~\eqref{eq:Cl_Bxtemp} and~\eqref{eq:Cl_tempxtemp} was verified to hold approximately for our simulations, despite the anisotropic LAT noise. (We note that the external-tracer simulations are masked by the LAT footprint but are otherwise statistically isotropic.) In an effort to reduce computational costs, the power spectrum obtained from the average of simulated lensing templates was therefore also used in our component separation pipeline to model cross-spectra between the template and observed $B$-modes from the different SAT channels. When working with real data, the template auto-spectrum may contain foreground residuals and should then be modelled separately from the cross-spectra.

\subsection{Galactic foreground model}
We now introduce the parametric model used to characterize SAT Galactic foregrounds and distinguish them from the CMB signal. The present analysis focuses on thermal dust emission and synchrotron radiation, as they are the two most prominent polarized sources over the scales of interest~\cite{krachmalnicoff_characterization_2016}.

\subsubsection{Baseline complexity}\label{subsec:d0s0}
In its simplest form, our model assumes foreground properties to be well-described by their sky average; the SEDs therefore depend on a restricted set of parameters that do not vary with position. 

The dominant contribution at low frequencies comes from synchrotron radiation, produced by the helical motion of high-energy cosmic ray electrons around Galactic magnetic field lines. Integrating the Larmor formula over a population of electrons with a power-law energy distribution, we obtain the SED $S_{\nu}^s=\left(\nu/\nu_0^s\right)^{\beta_s}$~\cite{rybicki_radiative_1979} in Rayleigh-Jeans units, where $\beta_s\approx-3$ and we fix $\nu_0^s=23\,\text{GHz}$. As a consequence of its negative spectral index, synchrotron radiation is subdominant above around $70\,\text{GHz}$ where thermal emission from dust grains heated by starlight becomes the most important foreground source. 

The dust component peaks in the far infrared and is polarized due to the alignment of dust particles with Galactic magnetic field lines, which arises from the action of radiative torques on irregularly shaped grains~\cite{kolokolova_grain_2015}. A modified blackbody function
\begin{equation}\label{eq:dust_sed}
I_{\nu}^d \propto \nu^{\beta_d}\frac{2h\nu^3}{c^2}
\frac{1}{\exp(h\nu/k_\text{B}T_d)-1}
\end{equation}
was empirically found to be a good approximation of its intensity spectrum~\cite{planck_collaboration_planck_2016}. The SED, $S_{\nu}^d=I_{\nu}^d/I_{\nu_0^d}^d$, then depends on the positive spectral index $\beta_d$ as well as on the interstellar dust temperature $T_d$ and the reference frequency $\nu_0^d$, which we fix at $353\,\text{GHz}$. \referee{Computing the sky average of the dust SED maps extracted from the Planck 2015 \texttt{Commander} analysis~\cite{planck_collaboration_planck_2016_foregrounds} yielded the reference values $\beta_d\approx1.54$ and $T_d\approx20\,\text{K}$, which are consistent with subsequent results from power-spectrum-based methods~\cite{planck_collaboration_planck_2020_dust}}.

Planck, WMAP and S-PASS observations have also shown that the foreground angular power spectra can be parametrized by power laws $C_l^c = A_c\left(l/l_0\right)^{\alpha_c-2}$, where $l_0 = 80$ and $c = d$ or $s$ for dust~\cite{planck_collaboration_planck_2020_dust} and synchrotron~\cite{fuskeland_constraints_2021} respectively. For each 
component, the slope of the angular power spectrum remains identical in all frequency channels, while its amplitude is multiplied by the product of the two relevant SEDs.

A dust-synchrotron correlation parameter $\epsilon_{ds}$ is also included in our model and contributes terms proportional to $C_l^{ds}=\epsilon_{ds}\sqrt{C_l^{d}C_l^{s}}$ to the power spectra. This brings the total number of free foreground parameters to seven: $\epsilon_{ds}$, $\beta_d$, $A_d$, $\alpha_d$, $\beta_s$, $A_s$ and $\alpha_s$.

\referee{Note that we do not include the Faraday rotation of primary $E$-modes into $B$-modes by Galactic or primordial magnetic fields in our foreground model. As the power spectrum of such $B$-modes scales as $\left(\nicefrac{\nu}{30\: \textrm{GHz}}\right)^{-4}$ and peaks at $l\sim1000$~\cite{de_cmb_2013}, this effect is expected to be negligible for the frequency bands and multipole range of the SO SATs.}

\subsubsection{Spatially-varying SEDs}\label{section:moments}
While the model described above is a useful approximation for pipeline validation purposes, it does not fully capture the complexity of realistic foreground emission. Indeed, the SEDs actually exhibit spatial variations due to inhomogeneous dust temperature and grain types as well as fluctuations in the cosmic ray energy distribution. Treating them as constants, especially over the large fractions of sky targeted by SO, has been shown to produce systematic residuals leading to significant biases in measurements of $r$ \cite{wolz_simons_2023}. 

In order to mitigate this effect, we apply the moment-expansion approach described in Ref.~\cite{azzoni_minimal_2021}. Assuming small spatial fluctuations $\delta\beta_c(\hat{\bm{n}})$ with respect to the sky average $\bar\beta_c$, we can expand the foreground contributions to second order at the map level\footnote{Variations in the dust temperature $T_d$ are almost degenerate with amplitude variations at the frequencies of interest, and contribute very little to SED variations. For this reason, we do not need to include the dust temperature in the moment expansion.}:
\begin{align}
\bm{m}_{\nu}(\hat{\bm{n}})&=\sum_c{\bm{T}_c(\hat{\bm{n}})S^c_{\nu}(\beta_c(\hat{\bm{n}}))} \nonumber \\
&=\sum_c\Bigl[\bm{T}_c(\hat{\bm{n}})\Bigl(\bar S^c_{\nu}+\delta\beta_c(\hat{\bm{n}})\partial_{\beta_c}\bar S^c_{\nu} \nonumber \\
&\mbox{} \hspace{0.15\textwidth} +\tfrac{1}{2}(\delta\beta_c)^2(\hat{\bm{n}})\partial_{\beta_c}^2\bar S^c_{\nu}\Bigr)\Bigr]. \label{eq:moments_map}
\end{align}
Here, $\bm{m}_{\nu}(\hat{\bm{n}})$ corresponds to the total foreground Stokes parameters at frequency $\nu$, $\bm{T}_c(\hat{\bm{n}})$ represents the amplitudes of the Stokes parameters of a given component at the reference frequency $\nu_0^c$ and the bars denote the SED or its derivatives evaluated at $\bar\beta_c$.

We can now propagate the expansion in Eq.~\eqref{eq:moments_map} to the cross-spectrum of the $B$-modes at frequencies $\nu$ and $\nu'$, using some simplifying assumptions justified in more detail in Ref.~\cite{azzoni_minimal_2021}. The leading-order term, 
\begin{equation}\label{eq:moments0x0}
C_{l,0}^{\nu\nu'}=\bar S_{\nu}^d\bar S_{\nu'}^dC_l^d+\bar S_{\nu}^s\bar S_{\nu'}^sC_l^s+\left(\bar S_{\nu}^d\bar S_{\nu'}^s+\bar S_{\nu}^s\bar S_{\nu'}^d\right)C_l^{ds},
\end{equation}
corresponds to the sky-average model presented in the previous subsection. While correlations between dust and synchrotron amplitudes are accounted for in Eq.~\eqref{eq:moments0x0}, we consider $\delta\beta_d$ and $\delta\beta_s$ to be independent. We further assume independence between $\bm{T}_c(\hat{\bm{n}})$ and $\delta\beta_c(\hat{\bm{n}})$.
Under these conditions, the first order terms vanish and the second-order contribution to the power spectrum is given by
\begin{multline}
C_{l,2}^{\nu\nu'}=\sum_{c}\bigg[\partial_{\beta_c}\bar S^c_{\nu}\partial_{\beta_c}\bar S^c_{\nu'}\sum_{l_1l_2}\tfrac{(2l_1+1)(2l_2+1)}{4\pi}\left(\begin{smallmatrix} l & l_1 & l_2 \\ 0 & 0 & 0 \end{smallmatrix}\right)^2 \\
\times C_{l_1}^cC_{l_2}^{\beta_c} + \frac{1}{2}\Bigl(\bar S^c_{\nu}\partial_{\beta_c}^2\bar S^c_{\nu'} + \bar S^c_{\nu'}\partial_{\beta_c}^2\bar S^c_{\nu}\Bigr)C_l^c \\ \times \sum_{l'} {\tfrac{2l'+1}{4\pi}}C_{l'}^{\beta_c}\bigg]. \label{eq:moments_spectrum}
\end{multline}

Further details of the derivation can be found in Ref.~\cite{azzoni_minimal_2021}; the important takeaway of Eq.~\eqref{eq:moments_spectrum} is that only one additional power spectrum per component, $C_l^{\beta_c}$, needs to be modeled in order to compute the effects of foreground spatial variability to second order in $\delta\beta_c(\hat{\bm{n}})$. Parametrizing it as a power law, $C_l^{\beta_c}=B_c\left(l/l_0\right)^{\gamma_c}$, leads us to extend our simple model by four variables. The full set of foreground parameters to be sampled is now given by $\epsilon_{ds}$, $\beta_d$, $A_d$, $\alpha_d$, $B_d$, $\gamma_d$, $\beta_s$, $A_s$, $\alpha_s$, $B_s$ and $\gamma_s$.

\subsection{Likelihood analysis}
In the likelihood analysis, we use the measured $B$-mode cross-spectra to infer the probability distributions of the foreground and cosmological parameters. 
Even if the underlying fields were Gaussian, dealing with the exact likelihood for the measured spectra in the presence of survey masking and inhomogeneous noise is intractable. We therefore adopt an approximate Gaussian likelihood for the measured spectra with a fixed covariance matrix. This approximation is expected to hold for large enough sky coverage and when enough multipoles are binned together and averaged, by virtue of the central-limit theorem. 

The Gaussian likelihood for the power spectra is
\begin{multline}\label{eq:gauss_likelihood}
    -2\ln{\mathcal{L}({\{\mathbf{X}_l\}|\{\hat{\mathbf{X}}_l\}})} \\
    \approx \left(\hat{\bm{X}}-\bm{X}\right)^T\mathbf{M}_{f}^{-1}\left(\hat{\bm{X}}-\bm{X}\right),
\end{multline}
where the vectors $\hat{\bm{X}}$ and $\bm{X}$ contain the measured cross-spectra and the model at all considered multipoles, respectively, and $\mathbf{M}_{f}$ is the fixed covariance matrix of the spectra. The covariance is precomputed from a set of simulations at fiducial parameter values and does not need to be recomputed as parameter space is explored, greatly reducing computational costs. A further advantage of working with a spectral likelihood is that non-Gaussianity of the fields (such as from lensing) can be dealt with approximately through the covariance matrix.

We also tested the approximate Hamimeche \& Lewis likelihood~\cite{hamimeche_likelihood_2008}, which similarly makes use of a fiducial covariance but additionally approximately captures the non-Gaussian shape of the likelihood. We compare results obtained with this likelihood and the Gaussian likelihood in Appendix~\ref{appendix_A}. As shown there in Fig.~\ref{fig:likelihood_comp}, the posterior distributions for the parameters are in excellent agreement; the Gaussian likelihood will therefore be used throughout this work as it leads to faster sampling. Note that the Gaussian approximation holds for SO due to the relatively large sky patch allowing to average over many modes. For experiments targeting smaller areas such as BICEP/Keck, non-Gaussian likelihood approximations, such as Hamimeche \& Lewis, must be used (especially at low $l$).

\section{Equivalence of cross-spectral and map-based delensing}\label{section:equivalence}

We now compare the expected performance of the power-spectrum-based approach presented above with that of map-based delensing, which will also likely be performed on future SO data. While directly subtracting the lensing $B$-mode template from the observed maps might seem more straightforward, cross-spectral methods have significant advantages for ground-based surveys. For example, complex filtering operations used to mitigate atmospheric disturbances are easier to take into account in harmonic space, and data splits can be used to provide an accurate estimate of the noise power spectrum (see Sec.~\ref{section:pipeline}). 

In an idealistic situation without such filtering, equivalent constraints on $r$ can be obtained with both methods. In order to demonstrate this, let us first consider a foreground-cleaned map from which we subtract the lensing template. The $B$-mode harmonic coefficients are given by $B_{lm}^{\textrm{del}}=B_{lm}+n_{lm}-B_{t,lm}$, where the subscript $t$ refers to the template. Assuming negligible foreground residuals, approximately Gaussian lensing $B$-modes and no correlations between the noise $n_{lm}$ and the other components, the power spectrum of this map then corresponds to
\begin{align}
C_l^{BB,\textrm{del}}&=C_l^{BB}-2C_l^{B_tB}+N_l+C_l^{B_tB_t}
\nonumber \\
&=C_l^{BB}-C_l^{B_tB_t}+N_l, \label{eq:map_del}
\end{align}
where we used the equality of Eqs~\eqref{eq:Cl_Bxtemp} and~\eqref{eq:Cl_tempxtemp} (which is verified later for our simulations in Fig.~\ref{fig:spectra}) to simplify the result. With the likelihood in Eq.~\eqref{eq:gaussian}, the Fisher information for $r$ at multipole $l$ is given by
\begin{equation}\label{eq:fisher_map}
\mathcal{I}_l(r)=\left\langle-\frac{\partial^2 \ln{\mathcal{L}}}{\partial r^2}\right\rangle=\frac{2l+1}{2}\left(\frac{\partial_r C_l^{BB}}{C_l^{BB}-C_l^{B_tB_t}+N_l}\right)^2.
\end{equation}
The error on $r$ is the same as in Eq.~\eqref{eq:sigma_r_fisher} with $C_l^{\textrm{lens}}$ replaced by its delensed counterpart.

Let us now consider the foreground-cleaned SAT map and the template separately. Again equating $C_l^{B_tB}$ and $C_l^{B_tB_t}$, the matrix $\mathbf{C}_l$ containing all auto- and cross-spectra becomes
\begin{equation}\label{eq:cross-spectra}
\mathbf{C}_l=\left(\begin{matrix}C_l^{BB}+N_l &  C_l^{B_tB_t} \\ C_l^{B_tB_t} & C_l^{B_tB_t}\end{matrix}\right).
\end{equation}
The Fisher information can be computed from the exact, full-sky likelihood for Gaussian fields, Eq.~\eqref{eq:gaussian_matrix}; using $\ln{|\mathbf{C}_l|}=\mathrm{Tr}\left(\ln{\mathbf{C}_l}\right)$, we obtain
\begin{equation}\label{eq:fisher_cross}
\mathcal{I}_l(r)=\frac{2l+1}{2}\mathrm{Tr}\left(\mathbf{C}_l^{-1}\partial_r\mathbf{C}_l\mathbf{C}_l^{-1}\partial_r\mathbf{C}_l\right).
\end{equation}
We can then substitute Eq.~\eqref{eq:cross-spectra} into this expression, where $C_l^{BB}$ is the only $r$-dependent element (neglecting small tensor contributions to the internal lensing estimators). As expected, the result is identical to Eq.~\eqref{eq:fisher_map} and we conclude that the map-based and the cross-spectral method lead to equivalent uncertainties on $r$. We verify this equivalence empirically in Sec.~\ref{section:results}.
Comparing constraints obtained with both techniques will therefore be a useful cross-check for future SO data.

Note that our reasoning was demonstrated on an idealistic full-sky situation (and with no foreground residuals) for simplicity; however, the argument still holds in the presence of a survey mask, provided that the same pixel weighting is used on all maps so that the cross-spectral likelihood takes as input all of the pseudo-spectra required to form the pseudo-spectrum of the map after map-based foreground cleaning and delensing.

\section{Optimal pixel weighting}\label{section:weighting}

For computational efficiency, we use local pixel weighting when measuring the power spectrum (i.e., we construct pseudo-$C_l$ estimates) rather than full inverse-variance weighting, which is non-local and requires inversion of large matrices. This local weighting scheme can be chosen to minimise the variance of the measured spectra (after deconvolution of the effects of the masking and weighting).
Building upon the results of Sec.~\ref{section:equivalence}, we construct optimal local pixel weights for a foreground-cleaned and delensed SAT map; these will be applied to all SAT channels as well as to the lensing template when computing pseudo-$C_l$ power spectrum estimates. A natural choice is the inverse-variance weighting
\begin{equation}\label{eq:weights}
w_i \propto \frac{1}{\sigma_{i,\textrm{res}}^2+\sigma_{i,\textrm{noise}}^2},
\end{equation}
where both the lensing residuals and the noise contribution (assumed to be constant over the pixel size) are taken into account.

This scheme can be heuristically motivated as follows. Assume that any statistical anisotropy due to the survey and the pixel weighting, are slowly varying compared to the scales of interest. We can then approximate the weighting as dividing up the sky into patches, large enough that the fluctuations are approximately uncorrelated between patches but small enough that the weights and statistical properties of the fluctuations can be treated as constant within each patch. Denote these patch weights by $w_i$, where $i$ labels the patch.
The observed $B$-mode power spectrum, corrected for the weighting, can then be approximated by a sum over patches:
\begin{equation}\label{eq:sum_patches}
\hat{C}_l\approx \left(\sum_i{w_i^2\tilde{C}_l^i}\right) / \langle w^2 \rangle_\text{patch} ,
\end{equation}
where $\tilde{C}_l^i$ is the pseudo-$C_l$ power spectrum in patch $i$ obtained with binary weighting (one in the patch and zero outside). The expected value of $\tilde{C}_l^i$ is approximately
\begin{equation}\label{eq:exp_localpower}
\langle \tilde{C}_l^i \rangle \approx f_{\textrm{sky},i} \left(C_l^i + N_l^i\right) ,
\end{equation}
where $C_l^i$ is the signal power spectrum in patch $i$, which may differ between patches due to statistical anisotropy (e.g., due to variation in delensing efficiency), $N_l^i$ is the local noise power spectrum there, and $f_{\textrm{sky},i}$ is the sky fraction of the patch. The normalisation factor in Eq.~\eqref{eq:sum_patches},
\begin{equation}
\langle w^2 \rangle_\text{patch} \equiv \sum_i w_i^2 f_{\textrm{sky},i} ,
\end{equation}
corrects for the effects of the weights and approximates the mask-deconvolution step in unbiased power-spectrum estimation.

The variance of $\hat{C}_l$ then corresponds to
\begin{equation}\label{eq:var_Cl}
\sigma^2(\hat{C}_l)\approx \left(\sum_i{w_i^4\sigma^2(\tilde{C}_l^i)}\right) / \langle w^2 \rangle_\text{patch}^2,
\end{equation}
where the variance of the $\tilde{C}_l^i$ is~\cite{knox_1995}
\begin{equation}
\sigma^2(\tilde{C}_l^i) \approx \frac{2}{(2l+1)f_{\textrm{sky},i}}\langle \tilde{C}_l^i \rangle^2.    
\end{equation}
Minimizing Eq.~\eqref{eq:var_Cl} with respect to $w_i$ leads to the weights
\begin{equation}
w_i \propto \frac{1}{C_l^i+N_l^i};
\end{equation}
as both the lensing $B$-modes and the noise power spectrum are approximately constant over the multipole range considered in our analysis ($30 \leq l \leq 300$), we indeed recover Eq.~\eqref{eq:weights}. 

The lensing residual term $\sigma^2_{i,\textrm{res}}$ is equivalent to $\left(5\,\mu \text{K-arcmin}\right)^2$ white noise multiplied by a factor $A_{\textrm{lens}}$ resulting from delensing. Due to the relative homogeneity of the LAT hit-count map (see Fig.~\ref{fig:hitmaps}), we take $A_\text{lens}$ to be uniform, computing it as\footnote{We give this more general form for completeness; in our analysis, the simpler form $A_\text{lens} = 1-\overline{C_l^{B_tB}/C_l^{BB}}$ is very nearly equivalent due to the equality of $C_l^{B_tB}$ and $C_l^{B_t B_t}$.}
\begin{equation}\label{eq:Alens}
A_{\textrm{lens}}=1-\overline{\left(\frac{(C_l^{B_tB})^2}{C_l^{BB}C_l^{B_tB_t}}\right)}, 
\end{equation}
where the bar refers to an average over $l$, on the whole area of overlap between the surveys with the SATs and the LATs~\cite{namikawa_simons_2022}. 
For future full-sky surveys such as LiteBIRD, which may be combined with a range of partially overlapping external tracers, the template properties might not be as uniform~\cite{namikawa_litebird_2023,belkner_cmb-s4_2023}. In this case, $A_{\textrm{lens}}$ could be treated as a spatially varying quantity. The noise term $\sigma^2_{i,\textrm{noise}}$ corresponds to the minimal white-noise variance in the deepest area of the map (hereafter referred to as the central white-noise variance $\sigma^2_{\textrm{white}}$) modulated by the relative number of hits:
\begin{equation}\label{eq:sigma_i_noise}
\sigma^2_{i,\textrm{noise}}=\frac{\sigma^2_{\textrm{white}}}{N^{\textrm{hit}}_i}.
\end{equation}

In the case of noise-dominated data, Eq.~\eqref{eq:weights} therefore reduces to the hit-count mask used in Ref.~\cite{wolz_simons_2023}; however, this is suboptimal for SO as the residual lensing variance surpasses the central white noise variance of the mid-frequency SAT channels. Similarly, the uniform weighting used in Ref.~\cite{namikawa_simons_2022} is only appropriate for areas in which noise inhomogeneities are negligible; mitigating the strongly increasing noise variance towards the outer regions of the map then requires aggressive apodization resulting in a loss of data. The strategy proposed here naturally transitions between these two extremes when moving from the edges to the center of the survey area.

With the lensing term characterized by Eq.~\eqref{eq:Alens}, the final step towards building our mask is the determination of the central noise variance $\sigma^2_{\textrm{white}}$ for a foreground-cleaned coaddition of SAT multi-frequency maps. For this purpose, we approximate the map-based foreground cleaning assuming known SEDs for all components.

In detail, 
we express the data $d_\nu(\hat{\bm{n}})$ at frequency $\nu$ as a sum of a noise contribution 
$n_\nu(\hat{\bm{n}})$ and three components $s_c(\hat{\bm{n}})$ (the CMB, dust and synchrotron) multiplied by their respective SEDs, which make up the mixing matrix $\mathbf{A}$ \cite{errard_robust_2016}:
\begin{equation}\label{eq:map_based_data}
d_\nu(\hat{\bm{n}})=\sum_c A_{\nu c}s_c(\hat{\bm{n}})+n_\nu(\hat{\bm{n}}).
\end{equation}
Assuming the noise to be Gaussian with covariance matrix $\mathbf{N}=\textrm{diag}\left(\sigma^2_{\rm{white},1},...,\sigma^2_{\rm{white},6}\right)$, where the central levels for individual channels are listed in Table~\ref{tab:SO}, we get the likelihood function
\begin{equation}\label{eq:map_based_likelihood}
-2\log{\mathcal{L}}=\left(\bm{d}-\mathbf{A}\bm{s}\right)^T\mathbf{N}^{-1}\left(\bm{d}-\mathbf{A}\bm{s}\right).
\end{equation}
We aim to compute the error on the element of $\bm{s}$ corresponding to the CMB~\cite{stompor_maximum_2009}. The associated Fisher information is given by
\begin{equation}\label{eq:map_based_fisher}
\mathcal{I}\left(s_{\textrm{CMB}}\right)=\left\langle-\frac{\partial^2 \ln{\mathcal{L}}}{\partial s_{\textrm{CMB}}^2}\right\rangle=\left(\mathbf{A}^T\mathbf{N}^{-1}\mathbf{A}\right)_{\text{CMB,CMB}},
\end{equation}
leading to the variance
\begin{equation}\label{eq:map_based_variance}
\sigma^2\left(s_{\textrm{CMB}}\right)=\sigma^2_{\textrm{white}}=\left(\mathbf{A}^T\mathbf{N}^{-1}\mathbf{A}\right)_{\text{CMB,CMB}}^{-1},
\end{equation}
which we can then substitute into Eq.~\eqref{eq:weights}. Considering the goal values presented in Table~\ref{tab:SO}, Eq.~\eqref{eq:map_based_variance} indicates a central white noise level of $2.5\,\mu\text{K-arcmin}$ for the foreground-cleaned map.

\section{Pipeline inputs and logistics}\label{section:pipeline}

\subsection{Simulations}\label{section:sims}

Full-sky CMB $Q$ and $U$ polarization maps are simulated as lensed Gaussian realizations of CMB power spectra computed with \texttt{CAMB}~\cite{camb_2000} for the Planck best-fit cosmological parameters listed in Ref.~\cite{calabrese_cosmological_2017}. One set of input simulations is produced with $r=0$ (no primordial $B$-modes) and the other one with $r=0.01$. The lensing operation is performed using the \texttt{pixell} package. These maps are generated at a high resolution ($N_{\textrm{side}}=2048$ in \texttt{HEALPix} \cite{gorski_healpix_2005}), then convolved with symmetrical Gaussian beams whose FWHMs are listed in Table~\ref{tab:SO}.

Our noise model accounts for detector white noise as well as a $1/f$ component related to timestream-level filtering (which mainly targets the mitigation of atmospheric noise), leading to the following angular power spectrum:
\begin{equation}\label{eq:noise_spectrum}
N_l = N_{\rm{white}}\left[1+\left(\frac{l}{l_{\rm{knee}}}\right)^{\alpha_{\rm{knee}}}\right].
\end{equation}
The amplitude $N_{\rm{white}}$ is adjusted for each instrument and frequency band in order for the central variance in the noise maps to match the values listed in Table~\ref{tab:SO}. Current or past experiments, such as QUIET \cite{quiet_collaboration_quiet_2013} near SO’s observing site and BICEP/Keck \cite{ogburn_iv_bicep2_2012} at the South Pole, have allowed to calibrate $l_{\rm{knee}}$ and $\alpha_{\rm{knee}}$ for the six SAT frequency channels (see Table~\ref{tab:SO}). For the LAT, we set $l_{\rm{knee}}=700$ and $\alpha_{\rm{knee}}=-1.4$ \cite{the_simons_observatory_collaboration_simons_2019}; these estimates based on ACT \cite{mallaby-kay_atacama_2021} noise levels are currently used for all LAT bands, however future work will aim to characterize the frequency dependency of these parameters. Noise maps are built by generating Gaussian realizations of the power spectrum in Eq.~\eqref{eq:noise_spectrum} (with $l_\text{max}=6144$ and $1536$ for the LAT and SAT, respectively), then multiplying each pixel $i$ by $1/\sqrt{N^{\textrm{hit}}_i}$, where $N^{\textrm{hit}}_i$ refers to the LAT or SAT normalised hit counts shown in Fig.~\ref{fig:hitmaps}.

\begin{table*}
    \begin{tabular}{c|cccc|cc} 
        \hline\hline
        & \multicolumn{4}{c |}{\thead{SAT}} & \multicolumn{2}{c}{LAT} \\
        \hline
        \thead{Frequency [GHz]} & FWHM [arcmin] & $\sigma_{\textrm{white},\nu}$ [$\mu{\rm K}$-arcmin] & $\ell_{\rm knee}$ & $\alpha_{\rm knee}$ & FWHM [arcmin] & $\sigma_{\textrm{white},\nu}$ [$\mu{\rm K}$-arcmin] \\
        \hline
        27 & 91 & 33 & 15 & -2.4 & 7.4 & 74 \\
        39 & 63 & 22 & 15 & -2.4 & 5.1 & 38 \\
        93 & 30 & 2.5 & 25 & -2.5 & 2.2 & 8.2 \\
        145 & 17 & 2.8 & 25 & -3.0 & 1.4 & 8.9 \\
        225 & 11 & 5.5 & 35 & -3.0 & 1.0 & 21 \\
        280 & 9 & 14 & 40 & -3.0 & 0.9 & 52 \\
        \hline
    \end{tabular}
    \caption{Noise and beam specifications used to produce the simulations in this work, corresponding to the goal/optimistic case of Refs~\cite{the_simons_observatory_collaboration_simons_2019} and \cite{wolz_simons_2023}. Note that the white-noise levels are given for the central area of the map assuming 5 years of observations, and are weighted according to the hit counts when generating the final noise maps. We use $l_{\rm{knee}}=700$ and $\alpha_{\rm{knee}}=-1.4$ for all LAT channels.}
    \label{tab:SO}
\end{table*}

The inputs of the internal lensing reconstruction stage simply consist of the CMB realizations described above convolved with LAT beams for three different frequency channels (93, 145 and 225 GHz), to which LAT-like noise is added. These maps do not contain any Galactic or extragalactic foregrounds, and therefore do not require point-source masks. 

In addition to LAT data, the template construction process relies on external lensing tracers. The present work follows Ref.~\cite{namikawa_simons_2022} and uses Gaussian simulations of the CIB as well as of a galaxy density field mimicking the forecasted LSST gold sample \cite{ivezic_lsst_2018}. The latter is split into six redshift bins with edges $z = \left[0, 0.5, 1, 2, 3, 4, 7\right]$. Relevant auto- and cross-spectra were computed in Ref.~\cite{yu_multitracer_2017} assuming linear galaxy bias. In order to ensure accurate correlation between the external tracers and the corresponding realization of the CMB lensing potential, we implement the algorithm described in Appendix F of Ref.~\cite{lizancos_delensing_2022}. The LAT binary mask derived from the hit counts in the left panel of Fig.~\ref{fig:hitmaps} is applied to all input maps, both internal and external, throughout the template construction step.

The next stage of the pipeline consists of extracting cosmological information from SAT polarization data. To simulate such observations, we use the same lensed CMB realizations as for the LAT-like maps, downgrade the resolution to $N_{\textrm{side}}=512$, convolve them with SAT beams in six frequency channels and add noise following Eq.~\eqref{eq:noise_spectrum} with the parameters listed in the left columns of Table~\ref{tab:SO}. Our choice to use the goal/optimistic noise levels from Ref.~\cite{the_simons_observatory_collaboration_simons_2019} is motivated by the recent approval of the SO:UK and SO:Japan projects, which are constructing three additional SATs; indeed, even if the nominal design of SO were to fall slightly short of these targets, our forecasts would still be conservative in light of the expected performance of the extended configuration. When analyzing real SAT observations, noise biases in the spectra will be mitigated by the use of data splits (see Sec.~\ref{sec:workflow}), corresponding to maps built from different time periods throughout the survey duration. In this paper, we use four splits (as was done for the ACT DR4 maps~\cite{aiola_atacama_2020}); this process is simulated by generating four uncorrelated noise realizations for each CMB map and multiplying the amplitudes from Table~\ref{tab:SO} by $\sqrt{N_{\rm{splits}}}=2$.

In order to assess the impact of Galactic foregrounds on delensing performance, our SAT-like simulations contain both dust and synchrotron emission. While past experiments have provided valuable information on these contaminants, the fluctuations of their SEDs across the sky have not been fully characterized yet. We therefore investigate three foreground models based on the \texttt{PySM} package templates \cite{thorne_python_2017} and including different degrees of complexity. All three models rely on dust and synchrotron amplitude maps established, respectively, by Planck at $353\,\text{GHz}$~\cite{planck_collaboration_planck_2016_foregrounds} and WMAP at $23\,\text{GHz}$~\cite{bennett_nine-year_2012}. The \texttt{Commander} component separation code also allowed to estimate the spatial distribution of the dust spectral index $\beta_d$ and temperature $T_d$ from the aforementioned Planck dataset. In the case of synchrotron radiation, spectral properties were inferred by combining WMAP data with the Haslam map at $408\,\text{GHz}$~\cite{haslam_408_1981,haslam_408-mhz_1982}. The \texttt{d1s1} foreground model used in our analysis is built upon this information.

A simpler model with uniform SEDs, referred to as \texttt{d0s0}, is derived from the \texttt{d1s1} templates by averaging the spectral parameters over the whole sky. The resulting values are $T_d=20\,\text{K}$, $\beta_d=1.54$ and $\beta_s=-3$. While it does not capture the full complexity of realistic foregrounds, this model is a very useful approximation for pipeline validation purposes as it allows the component separation process to be performed at a significantly lower computational cost. 

On the other hand, more recent data from S-PASS at $2.3\,\text{GHz}$~\cite{krachmalnicoff_s-pass_2018} seems to indicate that the synchrotron spectral index varies more significantly than in the original \texttt{PySM} template. A more complex model called \texttt{dmsm} is therefore constructed by rescaling the \texttt{d1s1} map of the fluctuations $\beta_s-\bar\beta_s$, using an amplification factor of 1.6. In this model, the spectral parameter templates for both dust and synchrotron are also smoothed at an angular resolution of $2\,\text{deg.}$ in order to mitigate instrumental noise residuals, which might have affected the \texttt{d1s1} maps. Gaussian fluctuations are then artificially added below this scale.

Foreground maps are convolved with the SO SAT-like Gaussian beams listed in Table~\ref{tab:SO} before being combined with the CMB and noise simulations, and are not subjected to bandpass integration. The choice of working with delta-function bandpasses simply aims to reduce computational costs here and does not significantly affect the results. (The analysis pipeline is also equipped to deal with realistic SO bandpasses.)

Finally, note that the foreground templates described above are informed by real data and are therefore only available in one realization. In order to evaluate the fiducial covariance matrix required to compute the likelihoods in Eqs~\eqref{eq:gauss_likelihood} and \eqref{eq:hnl_final}, 500 random Gaussian foreground simulations are generated using the angular power spectra mentioned in Section~\ref{subsec:d0s0} with reference parameters obtained from the \texttt{d0s0} dust and synchrotron maps. The best-fit power law parameters depend on the considered masking scheme, and thus exhibit slight variations with the value of $A_{\rm{lens}}$ in the optimal weights of Eq.~\eqref{eq:weights}. As the shifts observed in our case do not exceed 10\%, we use the same set of Gaussian simulations with and without delensing, and average the best-fit parameters obtained with $A_{\rm{lens}}=1$ and $A_{\rm{lens}}=0.35$. The resulting reference values are $A_d=44.6$, $\alpha_d=-0.16$, $A_s=1.07$ and $\alpha_s=-0.78$. The suitability of Gaussian simulations for covariance matrix estimation has been verified through $\chi^2$ tests in Appendix A of Ref.~\cite{wolz_simons_2023}; furthermore, Ref.~\cite{abril-cabezas_impact_2023} showed that modifying the covariance matrix to account for dust non-Gaussianity did not impact constraints on $r$ for an experiment with SO's characteristics.

\subsection{Workflow}\label{sec:workflow}

\begin{figure}[h]
    \centering
    \includegraphics[width=\linewidth, height=0.4\textheight]{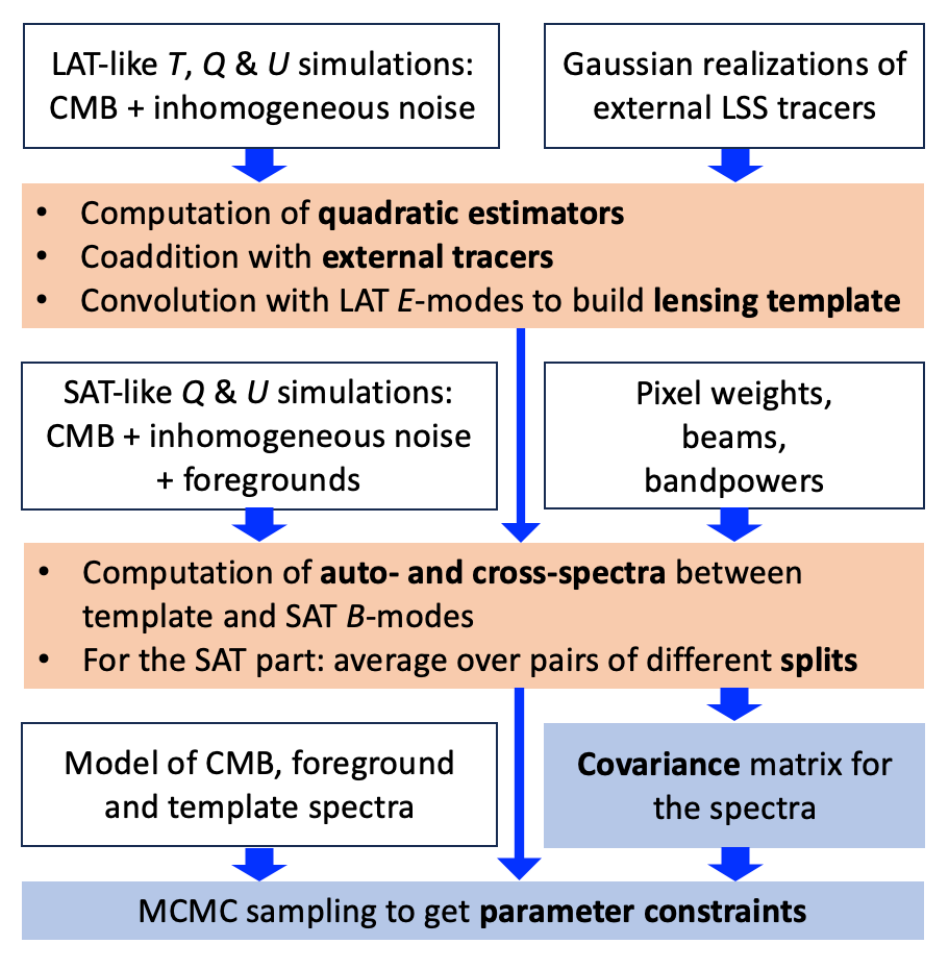}
    \vspace{-8pt}
    \caption{Flowchart of the three main pipeline stages: lensing template construction; computation of power spectra; and component separation. Inputs are listed in the white boxes. The same CMB realizations are used to generate the LAT and SAT maps, and the Gaussian realisations of external LSS tracers are appropriately correlated with the CMB lensing convergence. Orange boxes represent stages carried out on both the mock data and the 500 fiducial simulations, while the blue-colored operations are only applied to the mock data.}
    \label{fig:flowchart}
\end{figure}

The general working principle and successive stages of the analysis pipeline are summarized in Fig.~\ref{fig:flowchart}. When running the algorithm, one input map with either \texttt{d0s0}, \texttt{d1s1} or \texttt{dmsm} foregrounds is selected as mock data to run the parameter inference on, and the 500 simulations with Gaussian foregrounds are only used to compute the fiducial covariance matrix.

In the first stage, a binary mask based on the hit counts in Fig.~\ref{fig:hitmaps} and apodized by $5\,\text{deg.}$ is applied to all LAT-like simulations. The masked maps at 93, 145 and $225\,\text{GHz}$ are then converted to harmonic space and $E$-mode coefficients are combined with inverse-noise-variance weighting. Finally, these $E$-modes are diagonally Wiener filtered as indicated in Eq.~\eqref{eq:wiener}, where the coadded noise power spectrum $N_l^{EE}$ is obtained from noise-only simulations over the observed area.

The internal lensing convergence reconstruction process uses unfiltered $E$- and $B$-modes as well as temperature information extracted from the same LAT-like input maps. Diagonal inverse-variance filtering is directly incorporated in Eq.~\eqref{eq:kappa_estimator}, which is applied to four pairs of observed fields ($\Theta\Theta$, $\Theta E$, $EE$ and $EB$) in order to compute the quadratic estimators. 

Correlation coefficients between the minimum-variance combination of QEs and the external mass tracers, as well as between each tracer and the input lensing convergence are evaluated for all simulations. These values are averaged over the 500 CMB realizations and used in Eq.~\eqref{eq:kappa_comb} to coadd our internal reconstruction with the masked external-tracer maps. Convolving this combined estimator with the Wiener-filtered $E$-modes as indicated in Eq.~\eqref{eq:lensing_B} yields the harmonic coefficients of the lensing $B$-mode template, which we then convert into real-space $Q$ and $U$ maps. Note that the method used here to estimate the optimal coaddition of mass tracers will not be applicable to real data, for which only one realization is available and the true lensing convergence is not known. Instead, we will obtain the coefficients of Eq.~\eqref{eq:kappa_comb} by fitting theoretical models to the measured auto- and cross-spectra of our tracers. Uncertainties on these spectra were investigated in Ref.~\cite{namikawa_simons_2022} and found not to have a significant impact on $\sigma(r)$ at SO's sensitivity level.

\begin{figure*}[h!t]
    \centering
    \includegraphics[width=\linewidth, height=0.5\textheight]{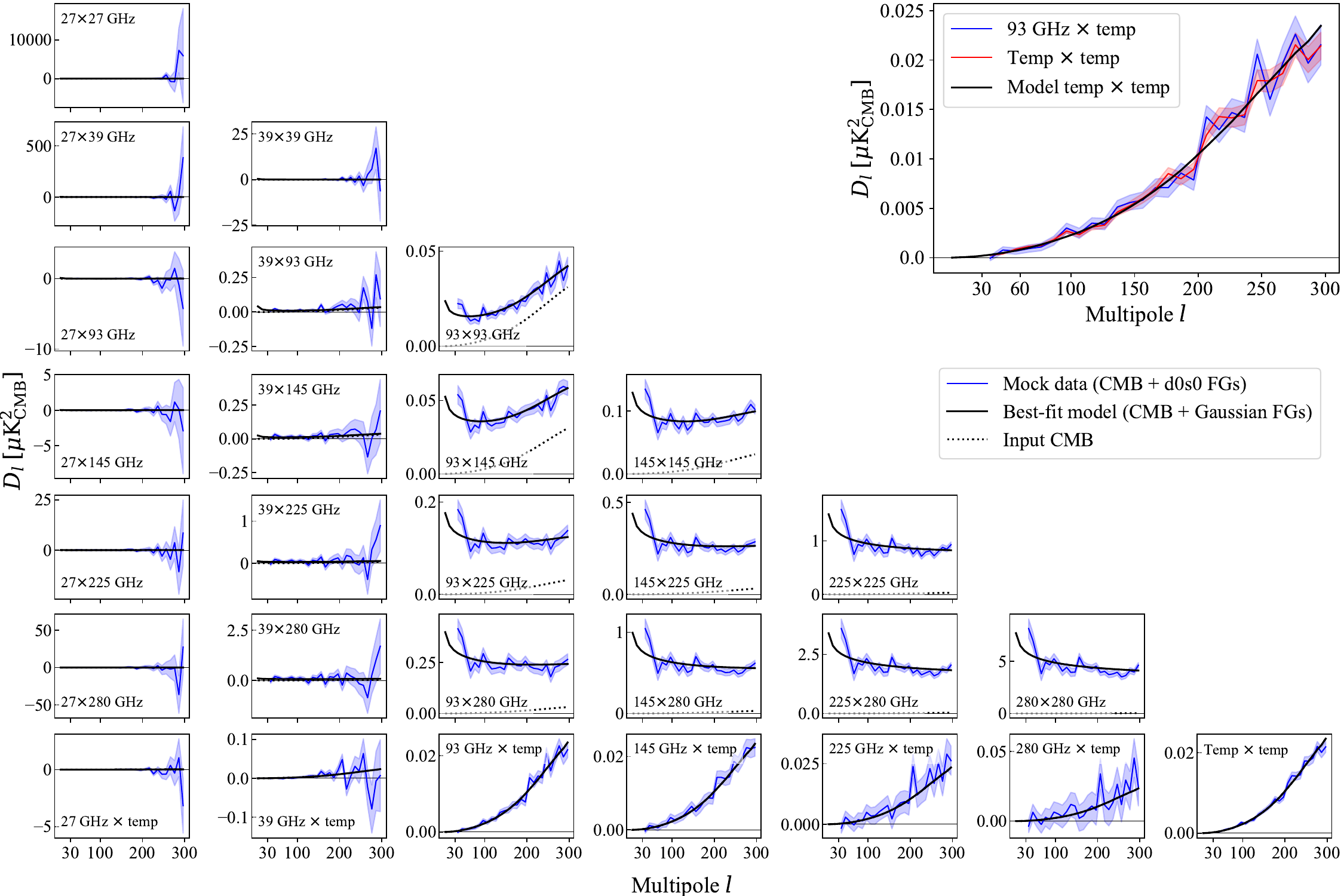}
    \vspace{-15pt}
    \caption{Power spectra computed for one realization of SAT $B$-modes in all frequency channels and the corresponding lensing template, with $D_l=l(l+1)C_l/2\pi$. The input maps include the CMB signal, SAT-like noise at goal/optimistic levels and \texttt{d0s0} foregrounds. The blue shaded areas represent the $1\sigma$ standard deviation calculated over 500 simulations with Gaussian foregrounds. The lensing template auto-spectrum and its cross-spectrum with SAT $B$-modes at $93\,\text{GHz}$ are highlighted in the top-right corner.}
    \label{fig:spectra}
\end{figure*}

\begin{table*}[h!tb]
    \begin{tabular}{ccccccccccccc} 
        \hline
        \hline
        \noalign{\smallskip}
        Parameter & $r$ & $\epsilon_{ds}$ & $\beta_d$ & $B_d$ & $\gamma_d$ & $A_d$ & $\alpha_d$ & $\beta_s$ & $B_s$ & $\gamma_s$ & $A_s$ & $\alpha_s$ \\
        \noalign{\smallskip}
        \hline
        \noalign{\smallskip}
        Prior & TH & TH & TH & TH & TH & TH & TH & G & TH & TH & TH & TH \\
        \noalign{\smallskip}
        Bounds & $[-0.1,0.1]$ & $[-1,1]$ & $[1.3, 1.8]$ & $[0, 10]$ & $[-6, -2]$ & $[20, 60]$ & $[-0.5, 0.5]$ & $-3 \pm 0.3$ & $[-5, 5]$ & $[-6, -2]$ & $[0, 4]$ & $[-3, 0]$ \\ 
        \noalign{\smallskip}
        \hline
    \end{tabular}
    \caption{Parameter priors for our MCMC analysis. The numbers listed here correspond to the center value and standard deviation in the Gaussian (G) case, or the lower and upper bounds for top-hat (TH) priors. Note that $B_d$, $B_s$, $\gamma_d$ and $\gamma_s$ only appear in the model when performing the moment expansion for spatially varying foregrounds.}
    \label{tab:priors}
\end{table*}

Throughout the second stage of the pipeline, both the SAT-like maps and the real-space $Q,U$ template are weighted according to Eq.~\eqref{eq:weights}, with the lensing and noise variance estimated as described in Sec.~\ref{section:weighting}.
Furthermore, the analysis region is restricted to the overlap between the SAT and the LAT survey areas, and an apodization length of $10\,\text{deg.}$ is applied. This additional smoothing is necessary for the $B$-mode purification process described below.

In the usual pseudo-$C_l$ formalism, the presence of such a mask/weighting leads to a mixing of the $E$- and $B$-modes as well as to the appearance of a coupling between different multipoles when computing auto- and cross-spectra between the observed SAT $B$-modes and the lensing template. Let us define the masked polarization vector $\bm{P}^w_{lm}=\left(E^w_{lm},B^w_{lm}\right)^{T}$, which contains the harmonic coefficients of the sky maps multiplied by the weights $w_i$ at the pixel level. The power spectrum estimator then corresponds to $\hat{\mathbf{C}}^w_l=\left(2l+1\right)^{-1}\sum_{m}{{\bm{P}^w_{lm}\bm{P}_{lm}^{w\dag}}}$, and the relation to the true spectrum (for statistically isotropic signals) is given by~\cite{hivon_master_2002}
\begin{equation}\label{eq:pseudo-Cl}
\textrm{vec}\left(\hat{\mathbf{C}}^w_l\right)=\sum_{l'}{\mathbf{M}^{22}_{ll'}\textrm{vec}\left(\hat{\mathbf{C}}_{l'}\right)},
\end{equation}
where the notation $\textrm{vec}\left(\mathbf{C}_l\right)$ refers to the vector of auto- and cross-power spectra.
The mode-coupling matrix $\mathbf{M}^{22}_{ll'}$, which can be computed directly from the pixel weights $w_i$, is generally not invertible due to the loss of information resulting from masking. This issue is resolved by grouping sets of $\Delta l$ multipoles into bandpowers and inverting the smaller binned form of $\mathbf{M}^{22}_{ll'}$. A bin width of $\Delta l=10$ is used in the present work.

As a consequence of the CMB $E$-mode power being significantly larger than that of $B$-modes, the aforementioned decoupling technique remains suboptimal. Indeed, $E$-mode leakage increases the variance of $\hat{C}_l^{BB}$, and this additional source of uncertainty needs to be mitigated at the map level \cite{lewis_analysis_2001}. This process, referred to as $B$-mode purification, is described in Ref.~\cite{smith_pseudo-c_ell_2006} and implemented in the \texttt{NaMaster} package \cite{alonso_unified_2019}. The Master algorithm~\cite{hivon_master_2002} also allows to perform beam deconvolution and invert the mode-coupling matrix $\mathbf{M}^{22}_{ll'}$, whose expression for purified pseudo $B$-modes is derived in Ref.~\cite{alonso_unified_2019}. 

After mitigating all mask-related artifacts, we still need to remove significant noise biases from our SAT $B$-mode auto-spectra (and, to a lesser extent, from cross-spectra between different frequency channels where noise may be at least partially correlated, for example due to atmospheric and instrumental systematic residuals). Our pipeline is designed to estimate the noise component directly from the data instead of relying on an instrumental model which might not be sufficiently precise. As briefly mentioned in Sec.~\ref{section:sims}, this is facilitated by the use of data splits, i.e., maps assembled from non-overlapping observations such that they do not share the same instrumental white noise realization. The final $\hat{C}_l^{BB}$ estimator between frequencies $\nu$ and $\nu'$ is then obtained by averaging the cross-spectra measured over all pairs of unequal splits $s_i, s_j$:
\begin{equation}\label{off_diag_coadd}
    \left(\hat{C}^{BB}_l\right)_{\nu\nu'}=\frac{1}{N_{\textrm{splits}}(N_{\textrm{splits}}-1)}\sum_{i\neq j}{\left(\hat{C}^{BB}_l\right)_{\nu\nu'}^{s_is_j}}.
\end{equation}

Due to the use of Wiener-filtered $E$-modes and lensing convergence tracers, noise-dominated multipoles are down-weighted in the power spectra \eqref{eq:Cl_Bxtemp} and \eqref{eq:Cl_tempxtemp} involving the lensing template, and no significant noise biases are observed for our simulations.
Our template is therefore built from full-survey maps instead of being subdivided into data splits, and its cross-spectrum with SAT $B$-modes at frequency $\nu$ is averaged as follows:
\begin{equation}\label{off_diag_coadd}
    \left(\hat{C}^{BB_t}_l\right)_{\nu}=\frac{1}{N_{\textrm{splits}}}\sum_{i}{\left(\hat{C}^{BB_t}_l\right)_{\nu}^{s_i}}.
\end{equation}
Template splits may, however, have to be introduced when working with real data, similar to their use in the recent ACT DR6 lensing analysis~\cite{qu_2024}.

Once all cross-spectra are calculated for the 500 fiducial model simulations, they are used to compute the sample covariance matrix $\mathbf{M}_{f}$ required for the likelihood in Eq.~\eqref{eq:gauss_likelihood}. This matrix is conditioned to be block-diagonal, with all correlations between multipoles separated by more than the bin width $\Delta l$ being set to zero. Such a precaution is necessary in order to remove spurious off-diagonal elements related to Monte-Carlo noise, a consequence of estimating the covariance from a finite number of simulations, which has been shown to result in bias and underestimated uncertainties when inferring $r$~\cite{beck_bias_2022}.

\begin{figure*}[h!bt] 
    \centering
    \includegraphics[width=\linewidth]{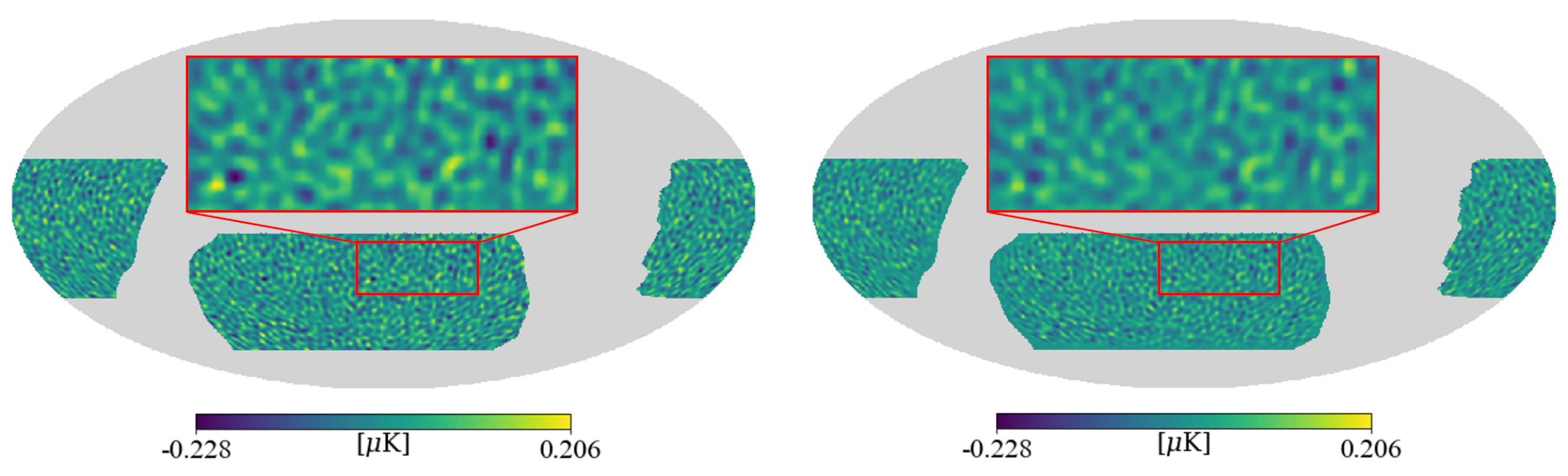}
    \vspace{-15pt}
    \caption{\textit{Left:} lensing $B$-modes in one of the input CMB maps, projected as a scalar field onto the overlap area between the SAT and LAT masks. \textit{Right:} corresponding lensing $B$-mode template projected in the same way. In both panels, we only show multipoles between $20 \leq l \leq 128$ and include a zoomed-in region to highlight the visual correlation between the two maps.}
    \label{fig:temp_maps}
\end{figure*}

Example power spectra are displayed in Fig.~\ref{fig:spectra} for one specific CMB realization with \texttt{d0s0} foregrounds, using optimal pixel weights. The $1\sigma$ standard deviation on the spectra, represented by the blue shaded areas, is given by the square root of the diagonal covariance elements. In the two left-hand columns, corresponding to the 23 and $39\,\text{GHz}$ bands, large noise fluctuations appear at high multipoles. This amplified variance is a consequence of beam deconvolution in the frequency channels with the lowest resolution. The shape of the lensing $B$-mode spectrum (dotted line) is mostly recognizable at 93 and $145\,\text{GHz}$, while the two highest frequency bands are heavily dominated by dust emission. For $l>40$, all panels display a good agreement between the measured power spectra and the fiducial model described in Sec.~\ref{subsec:d0s0} (with the best-fit parameters cited at the end of Sec.~\ref{section:sims}). The steepening of the power spectra on large scales, which is not observed in Ref.~\cite{wolz_simons_2023}, is a consequence of the less conservative survey mask used in the present work.
With narrower Galactic cuts, the foregrounds contained in our simulations are brighter and more complex, deviating from our simple power law fiducial model. A curvature term may be added to the exponent in future work in order to obtain a better fit. Alternatively, 
assumptions regarding the shape of the power spectrum may be removed by using an $l$-wise parametrization, or by working with empirically determined spectral templates. When working with real data, different combinations of Galactic cuts and model parametrizations will be tested in order to determine the optimal analysis settings. While a more conservative mask may end up being selected at this stage, we intentionally keep the wider survey area here in order to demonstrate the proper functioning of our pipeline in the presence of complex foreground features.

In the subplots in Fig.~\ref{fig:spectra} involving the lensing template, the black continuous line represents the mean of $\hat{C}_l^{B_tB_t}$ over the 500 CMB realizations. This closely matches the mean of the cross-spectra of the template with SAT $B$-modes $\hat{C}_l^{BB_t}$, thus explicitly verifying that the equivalence discussed in Sec.~\ref{section:equivalence} still holds in the presence of inhomogeneous LAT noise. This conclusion is further supported by the highlighted subplot in the top-right corner of Fig.~\ref{fig:spectra}, where correlated fluctuations are visible between $\hat{C}_l^{B_t B_t}$ and $\hat{C}_l^{B B_t}$ at $93\,\text{GHz}$.
We therefore use the mean of $\hat{C}_l^{B_t B_t}$, computed for the fiducial simulations using the apodized optimal weighting mask, as our model for all spectra involving the lensing template.

In the final stage of the pipeline, the measured cross-spectra, their covariance matrix and our parametric model are substituted into the likelihood function, which can either be the Gaussian likelihood in Eq.~\eqref{eq:gauss_likelihood} or the Hamimeche \& Lewis approximation~\eqref{eq:hnl_final}. Note that we only select multipoles between 30 and 300; this range targets the scales at which primordial $B$-modes are expected to be the most significant while preventing our analysis from being affected by effective large-scale noise induced by timestream filtering. A Markov Chain Monte Carlo (MCMC) algorithm implemented in the \texttt{emcee} package is then run for 8000 iterations in order to sample the posterior distributions of $r$ and our foreground model parameters. The number of walkers is set to 24 for \texttt{d0s0} input simulations, and increased to 48 for the more complex \texttt{d1s1} and \texttt{dmsm} models. In order to ensure convergence, we estimate the integrated autocorrelation time (IAT) for $r$ and obtain 81 for the baseline analysis and 191 for the moment expansion method, thus confirming that the length of our chains is sufficient (between 40 and 100 times the IAT). Table~\ref{tab:priors} lists the priors used for parameter estimation; most of them follow Ref.~\cite{wolz_simons_2023}, with the exception of $\beta_d$ for which the Gaussian prior is replaced by a top-hat. This choice, physically motivated by the $\beta_d$ map extracted from Planck data and shown in Ref.~\cite{zonca_python_2021}, was found to lead to faster convergence in the presence of complex foregrounds. It is worth mentioning that the prior on $r$ will be restricted to positive values when analyzing real data, but is deliberately kept open for now in order to check for potential biases.

\section{Results and discussion}\label{section:results}
\subsection{Lensing template}\label{section:results_temp}

The lensing $B$-mode template obtained from one realisation of our LAT-like simulations and the associated (correlated) realisation of the external LSS tracers is displayed in the right panel of Fig.~\ref{fig:temp_maps}, and compared to the input lensing $B$-modes in the same CMB realization. Both subplots only contain a restricted set of large-scale multipoles ($20 \leq l \leq 128$) for legibility purposes, and show the region of overlap between the LAT and SAT surveys. The mask is different from the one presented in Ref.~\cite{namikawa_simons_2022} as updated versions of the hit-count maps are used here. In particular, Galactic cuts are less conservative in the present work than in Ref.~\cite{namikawa_simons_2022} in order to maximize the analysis area. 

The correlation between the template and the original $B$-modes is clearly visible in Fig.~\ref{fig:temp_maps}, especially in the zoomed-in rectangles. A certain degree of attenuation is noticeable when comparing the two panels of the figure. This is a direct consequence of the use of Wiener-filtered $E$-modes and lensing convergence tracers in Eq.~\eqref{eq:lensing_B}; indeed, Eqs~\eqref{eq:wiener} and \eqref{eq:kappa_comb} imply that any noise in the LAT maps results in a decreased amplitude for $\hat{E}_{lm}^{\rm{WF}}$ and $\hat{\kappa}_{LM}^{\textrm{comb}}$.

\begin{figure}[h!]
    \centering
    \includegraphics[width=\linewidth]{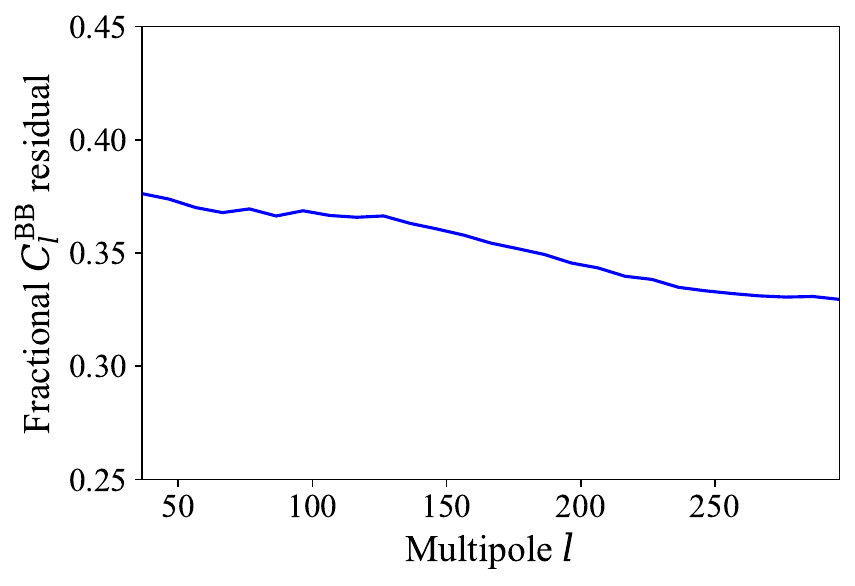}
    \vspace{-15pt}
    \caption{Ratio between the residual and input lensing $B$-mode power, obtained as in Eq.~\eqref{eq:Alens} but without averaging over $l$, in the overlap area between the SAT and the LAT surveys. Using LAT-only $E$-modes combined with a multitracer $\kappa$ estimator and considering the goal LAT noise level, the delensing efficiency of our template reaches approximately 65\%.}
    \label{fig:Alens}
\end{figure}

The fractional lensing $B$-mode power residual after map-level delensing with our template is displayed in Fig.~\ref{fig:Alens}. Considering the minimum-variance delensed $B$-modes\footnote{We give the general form here, involving the Wiener-filtered template, although in our analysis the filter is very close to unity since $C_l^{BB_t} \approx C_l^{B_t B_t}$.}~\cite{namikawa_simons_2022}
\begin{equation}
B_{lm}^{\textrm{del}}=B_{lm}-\frac{C_l^{BB_t}}{C_l^{B_tB_t}}B_{t,lm},
\end{equation}
the average of the ratio $C_l^{BB,\textrm{del}}/C_l^{BB}$ over $l$ corresponds to the $A_{\textrm{lens}}$ parameter defined in Eq.~\eqref{eq:Alens}. In practice, we evaluate this quantity by computing the pseudo-$C_l$ of the noise-free input lensing $B$-modes ($C_l^{BB}$) and of the lensing template ($C_l^{B_tB_t}$), as well as their cross-spectrum $C_l^{BB_t}$, using the binary SAT/LAT overlap mask with an apodization length of $5\,\text{deg}$.

\begin{figure*}[h!bt]
    \centering
    \begin{minipage}{.48\textwidth}
        \centering
        \includegraphics[width=\linewidth, height=0.25\textheight]{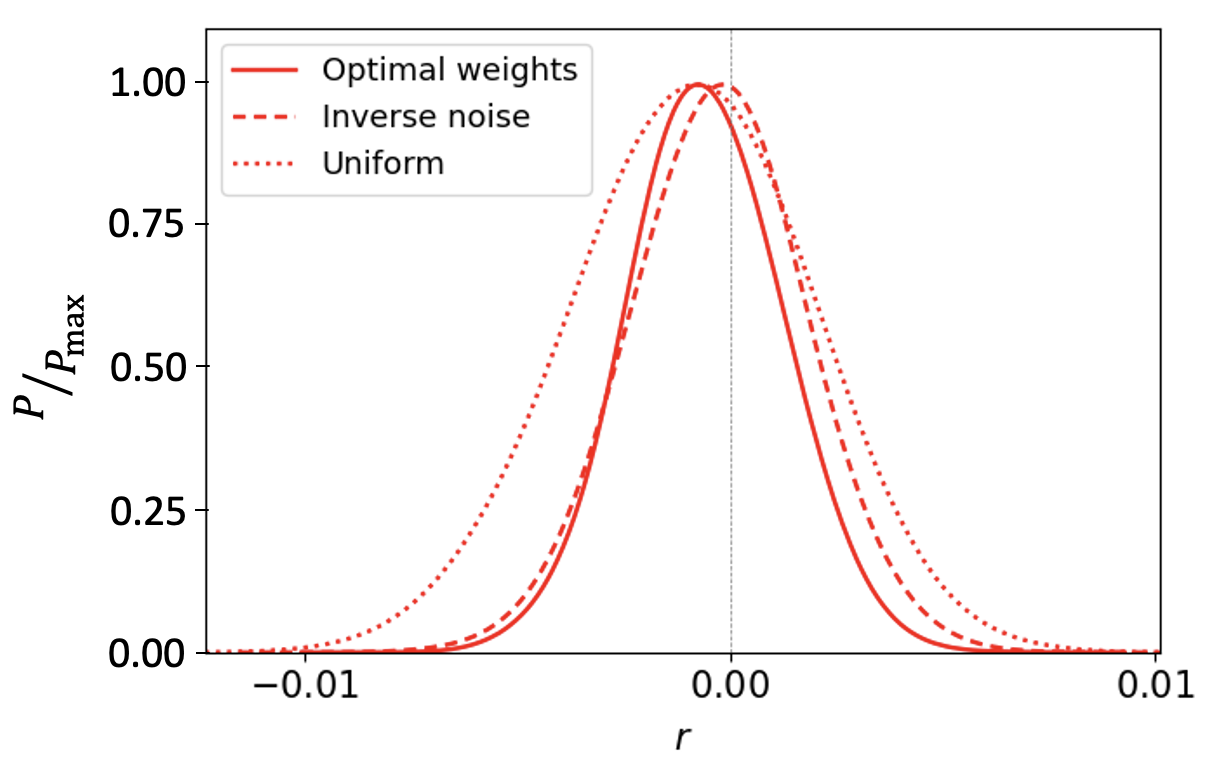}
    \end{minipage}%
    \hfill
    \begin{minipage}{0.48\textwidth}
        \centering
        \includegraphics[width=\linewidth, height=0.25\textheight]{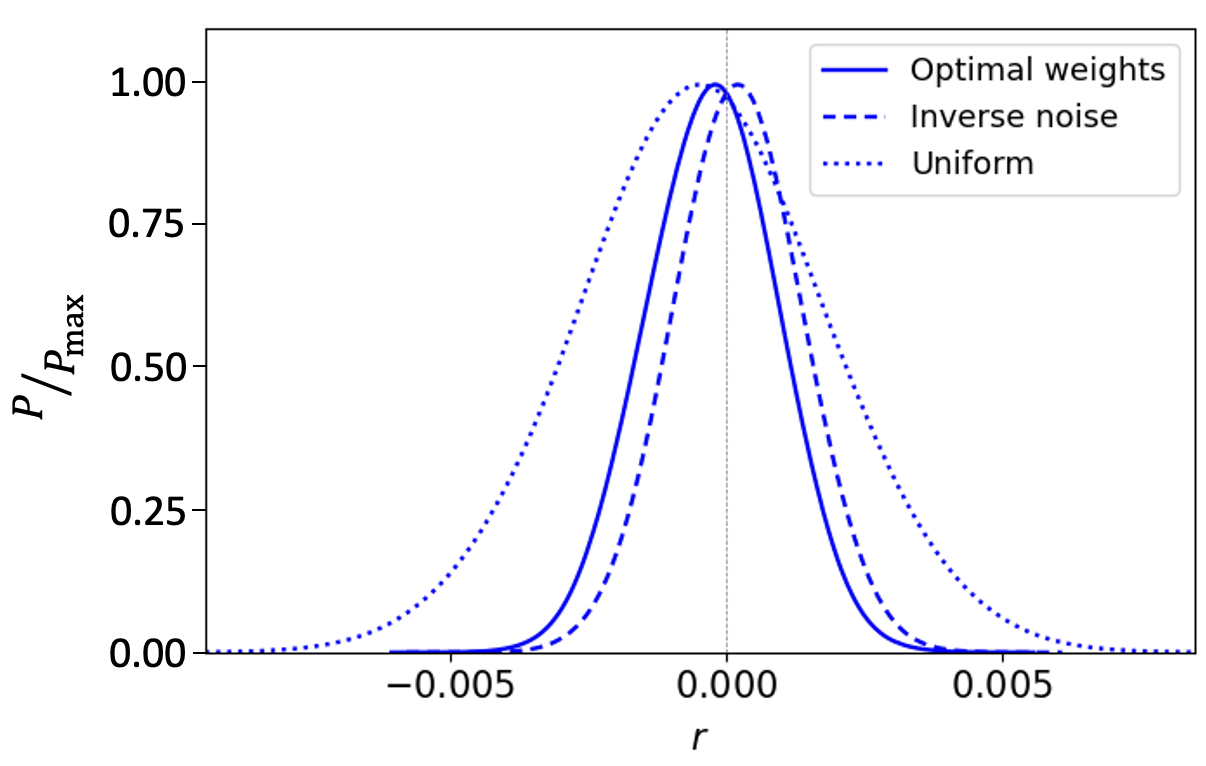}
    \end{minipage}
    \vspace{-5pt}
    \caption{Marginalized posterior distributions for the tensor-to-scalar ratio $r$ obtained by sampling the Gaussian likelihood Eq.~\eqref{eq:gauss_likelihood} for a single input realization. The input simulation has $r=0$ and \texttt{d0s0} foregrounds, and cross-spectra are computed using uniform (dotted lines), hit-count (dashed lines) and optimal (solid lines) pixel weights, respectively. \textit{Left:} no delensing is applied. \textit{Right:} cross-spectral delensing is performed using the template characterized in Sec.~\ref{section:results_temp}.}
    \label{fig:mask_comp}
\end{figure*}

Figure~\ref{fig:Alens} illustrates the slight dependence of $C_l^{\tilde{B}\tilde{B},\textrm{del}}/C_l^{\tilde{B}\tilde{B}}$ on angular scale, and is consistent with the results obtained in Ref.~\cite{namikawa_simons_2022} for LAT-only $E$-modes with the previous version of the survey masks. The fraction of lensing $B$-mode power remaining after delensing averages out at 35\%. As shown in Fig.~9 of Ref.~\cite{namikawa_simons_2022} for an idealistic foreground-free situation, including SAT $E$-modes in the template construction stage may allow to bring this ratio further down towards 30\%. However, applying this operation to real data would also require a careful treatment of foreground residuals, which we leave for future work.

\subsection{Pixel weighting}

\begin{figure*}[h!t]
    \centering
    \includegraphics[width=\linewidth]{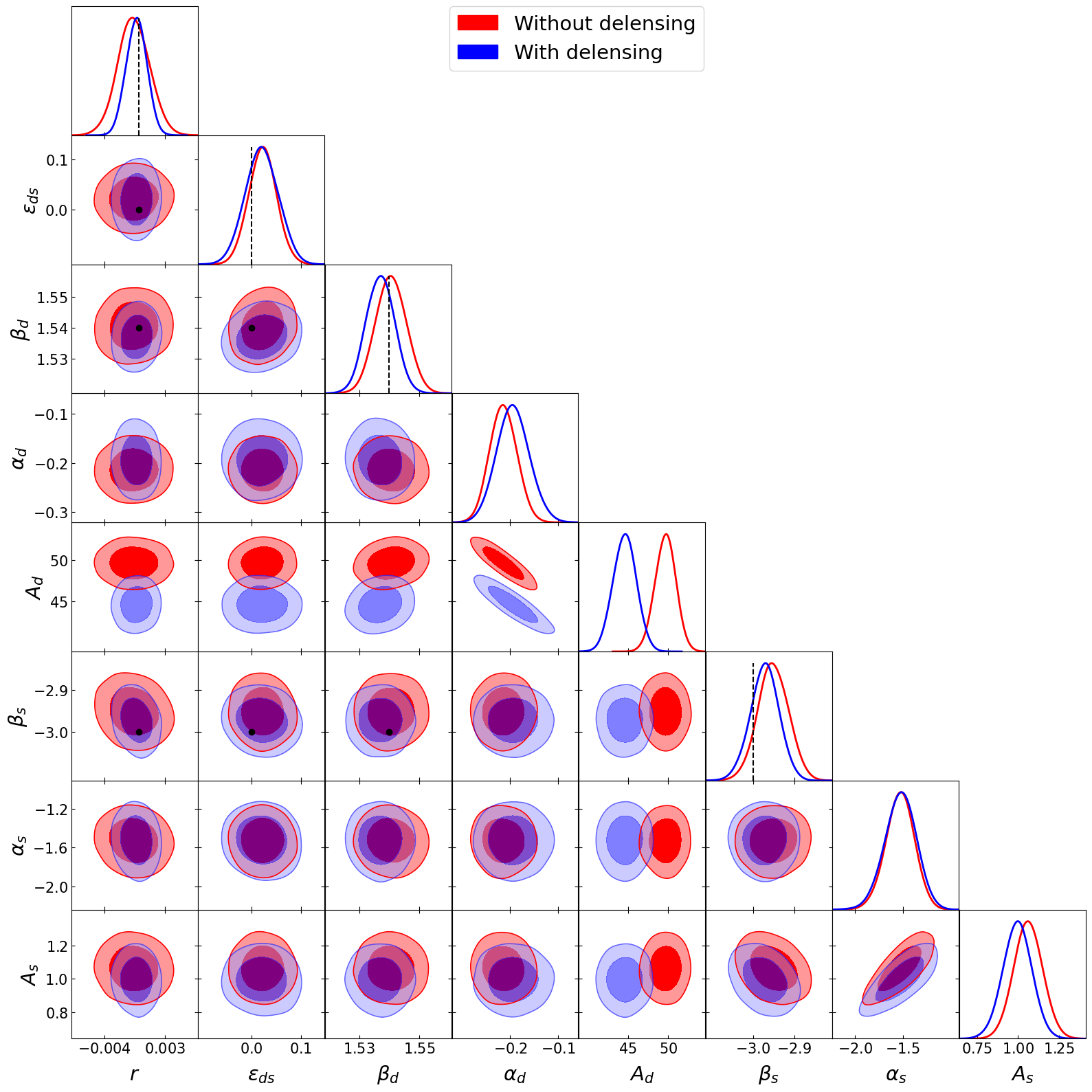}
    \vspace{-15pt}
    \caption{Triangle plot displaying the posterior distributions obtained for all model parameters with (blue) and without (red) delensing, using one input map realization with \texttt{d0s0} foregrounds. The moment expansion is not performed here. For $r$, $\epsilon_{ds}$, $\beta_d$ and $\beta_s$, the dashed lines and dots represent the true values used to generate our simulations. The inferred parameters for the foreground angular power spectra depend on pixel weights (as visualized by the shift of the $A_d$ posterior), therefore we do not plot ground truths for $A_d$, $\alpha_d$, $A_s$ and $\alpha_s$.}
    \label{fig:triangle_comp}
\end{figure*}

We now present a series of tests performed in order to check the proper functioning of our pipeline and validate the optimal pixel weights discussed in Sec.~\ref{section:weighting}. The \texttt{d0s0} foreground model is used in our input maps throughout this section, as it is simple enough to incur low computational costs but realistic enough to provide informative results. We use 500 simulations containing the CMB, SAT-like noise and Gaussian foregrounds to evaluate the covariance matrix in each variant of the analysis.

Our aim is to verify that the pixel weights in Eq.~\eqref{eq:weights} do lead to tighter cosmological constraints than the uniform or inverse-noise-variance ($\propto N^{\textrm{hit}}_i$) weights used in Refs~\cite{namikawa_simons_2022} and~\cite{wolz_simons_2023}, respectively. This is done by analysing the same input map with different weighting schemes and recomputing all cross-spectra as well as the corresponding covariance matrix each time. The apodization length is set to $10\,\text{deg.}$ for the optimized and hit-count weights; we increase it to $25\,\text{deg.}$ for the uniform weights which do not otherwise fall off smoothly near the edge of the survey area. We use $A_{\rm{lens}}=1$ to compute the optimal weights without delensing, and change it to $0.35$ when delensing is performed.

The marginal posterior distributions for $r$ are obtained in each case for one realization are displayed in Fig.~\ref{fig:mask_comp}, confirming that our optimal weights (solid lines) indeed perform better than the inverse-noise and uniform cases. The uniform weighting is clearly suboptimal, yielding the MCMC standard deviation $\sigma(r)=2.9\times 10^{-3}$ before delensing and $2.3\times 10^{-3}$ after. The larger statistical errors and the modest improvement from delensing (only 21\%) are a consequence of the high effective noise level in the outer regions of this mask, which leads $N_l$ to dominate in Eq.~\eqref{eq:sigma_r_fisher}. On the other hand, the distributions obtained with the hit-count mask and the optimal pixel weights are very similar. Their respective standard deviations are $\sigma(r)=2.1\times 10^{-3}$ and $1.9\times 10^{-3}$ with the baseline analysis; once delensing is applied, both values decrease to about $1.2\times 10^{-3}$. At SO's nominal sensitivity,
the difference between the two methods is therefore not very significant. As expected, it is more apparent without delensing and practically invisible in the $A_{\rm{lens}}=0.35$ case, where instrumental noise still dominates over lensing residuals. However, our new weighting scheme will become increasingly relevant for future experiments with lower noise levels (for which the lensing $B$-mode variance in Eq.~\ref{eq:weights} will be comparatively more important), and in particular for LiteBIRD where the residual lensing $B$-modes will vary across the sky~\cite{namikawa_litebird_2023}.

The derivation presented in Sec.~\ref{section:weighting} assumes instrumental noise to be well-described by a white power spectrum. To remove this assumption, we also investigated an alternative technique inspired by the hybrid $C_l$ estimator described in Ref.~\cite{efstathiou_myths_2004}, which combines power spectra obtained with uniform and hit-count weights. Due to issues related to its very high computational costs, this procedure is not yet suitable for use in SO data analysis. A more detailed discussion of the hybrid-weighting method and of the problems arising when testing it on SO-like simulations can be found in Appendix~\ref{appendix_B}.
The optimised weights in Eq.~\eqref{eq:weights} are systematically used for all results that follow in the main text.

All final outputs of our analysis for a given input simulation are shown in Fig.~\ref{fig:triangle_comp}, where posterior distributions for $r$ and the foreground parameters can be seen with and without delensing. As the \texttt{d0s0} foreground map used here does not include SED spatial variability, we consider the simplest form of our parametric model and do not perform the moment expansion. The $r$ distribution visibly tightens as a result of delensing, with $\sigma(r)$ dropping from $1.9\times 10^{-3}$ to $1.2\times 10^{-3}$ (37\% improvement). 

The only foreground parameter whose posterior significantly changes after delensing is the dust amplitude $A_d$; this is easily explained by the dependence of the weights in Eq.~\eqref{eq:weights} on $A_{\textrm{lens}}$. Indeed, without delensing, the optimized mask is slightly closer to uniform than with $A_{\textrm{lens}}=0.35$. Pixels near the Galactic plane are then weighted more heavily, resulting in a higher effective dust amplitude. The \texttt{d0s0} best-fit estimates are respectively $A_d=47.2$ without delensing and $42.1$ with, averaging to $44.6$. The observed shift of the $A_d$ maximum posterior by about 5 units is therefore consistent with our expectations. The synchrotron amplitude and the power law exponents exhibit less significant shifts; in all cases, priors are chosen wide enough to hold for both values of $A_{\rm{lens}}$.

\renewcommand{\arraystretch}{1.6}
\begin{table*}[h!tb]
        \begin{tabularx}{0.95\linewidth}{YYYYYYYY} 
            \hline
            \hline
            \multicolumn{1}{c}{} & \multicolumn{3}{c}{\thead{\textbf{Without delensing}}} & \multicolumn{3}{c}{\thead{\textbf{With delensing}}} & \multicolumn{1}{c}{\thead{$\bm{A}_{\textbf{lens}}\bm{=0.35}$}}\\
            \hline
            \textbf{FG model} & Mean of $\hat{r}$ & Mean of $\sigma(r)$ & SD of $\hat{r}$ & Mean of $\hat{r}$ & Mean of $\sigma(r)$ & SD of $\hat{r}$ & $\sigma(r)$ \\
            & ($\times10^3$) & ($\times10^3$) & ($\times10^3$) & ($\times10^3$) & ($\times10^3$) & ($\times10^3$) & ($\times10^3$) \\
            \hline
            \texttt{d0s0} & 0.30 & 1.9 & 1.7 & 0.20 & 1.2 & 1.1 & 1.2 \\
            \texttt{d1s1} & -0.35 & 3.2 & 2.6 & -0.53 & 2.2 & 2.1 & 2.4 \\
            \texttt{dmsm} & -0.73 & 3.0 & 2.5 & -0.34 & 2.2 & 2.0 & 2.3 \\
            \hline
        \end{tabularx}
        \caption{Tensor-to-scalar ratio statistics averaged over 100 simulations for three increasingly complex foreground models and $r=0$. The moment expansion of the SEDs is performed for \texttt{d1s1} and \texttt{dmsm}. The quantities $\hat{r}$ and $\sigma(r)$ refer to the maximum of the $r$ posterior distribution and the SD of the corresponding MCMC samples, respectively. Values in the last column are obtained by decreasing the lensing $B$-mode power in the input simulations and running the original pipeline (without cross-spectral delensing).}
        \label{tab:del_performance}
\end{table*}

Note that the $\alpha_s$ posterior in Fig.~\ref{fig:triangle_comp} peaks significantly lower than the reference value mentioned in Sec.~\ref{section:sims} ($\alpha_s=-0.78$); this is due to the fact that the synchrotron angular power spectrum is not a perfect power law and has a steeper slope on large scales. As the noise level is strongly amplified by beam deconvolution
above $l\approx200$ at 27 and $39\,\text{GHz}$, where synchrotron emission dominates, the SATs are mostly sensitive to $\alpha_s$ at low multipoles. This results in our pipeline inferring a steeper power-law index than the best-fit value obtained from the \texttt{s0}-only map over the full $30\leq l \leq 300$ range (which is how the fiducial value is obtained).
Figure~\ref{fig:triangle_comp} does not indicate any significant degeneracy between $r$ and $\alpha_s$; our cosmological constraints are therefore unlikely to be affected, however a more complex synchrotron model including a curved power-law index may be used in future work.

\subsection{Delensing performance with realistic foregrounds}\label{sec:results_delensing}

Having demonstrated the pipeline on the simplest of our sky models, we now assess the impact of delensing in the presence of more complex foregrounds as well as nonzero tensor modes. In each case considered, we also check for biases and verify the robustness of our statistics by computing the maximum-posterior estimate $\hat{r}$ and the width of the distribution $\sigma(r)$ for 100 different input maps. These simulations correspond to random realizations of the CMB and noise, with a fixed PySM foreground template. For \texttt{d1s1} and \texttt{dmsm}, we perform the moment expansion described in Sec.~\ref{section:moments} in order to account for the spatially varying SEDs. Following Ref.~\cite{wolz_simons_2023}, we use the same fiducial covariance matrix when changing the foreground model, however it must be modified to account for the variance of tensor modes in the $r=0.01$ case. 

As a first sanity check, we verify (using our $r=0$ simulations) that the standard deviation (SD) of the 100 best-fit estimates $\hat{r}$ is broadly consistent with the average $\sigma(r)$ obtained from the SD of the MCMC samples for each realization (see Table~\ref{tab:del_performance}). Note that the 100 simulations used as mock data contain the same foreground map; we therefore expect the SD estimated from this set of inputs to be lower than the MCMC $\sigma(r)$ value determined by the covariance matrix, which accounts for the effects of foreground variance. For this reason, we choose the latter as the preferred statistic (it is also the only one we will have access to when working with real data). 

We then check that $\sigma(r)$ does not scatter significantly: its SD over the 100 simulations is of the order $10^{-5}$ for \texttt{d0s0} and $10^{-4}$ for \texttt{d1s1} and \texttt{dmsm}. This result is expected, as error bars are mostly determined by the covariance matrix (which remains unchanged) and should not strongly depend on the input realization. Finally, we compare the SD of the mean of the posterior $r_{\textrm{mean}}$ to that of $\hat{r}$. With a difference of less than 2.5\% in all foreground cases, the two values are in good agreement both with and without delensing, as expected for Gaussian posteriors.

\begin{figure}[h!]
    \centering
    \includegraphics[width=\linewidth]{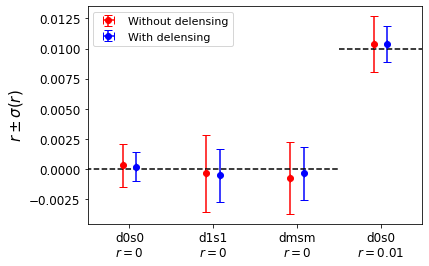}
    \vspace{-15pt}
    \caption{Comparison of the mean of $\hat{r}$ and the MCMC standard deviation $\sigma(r)$ (represented by the error bars) with and without delensing. We average over 100 simulations for each case. The moment-expansion method is used for \texttt{d1s1} and \texttt{dmsm} foregrounds, resulting in larger error bars. The presence of tensor modes is successfully detected by the pipeline and also leads to increased uncertainties. Delensing causes the constraints to tighten by 27\% to 37\% depending on the characteristics of the input maps.}
    \label{fig:del_performance}
\end{figure}

The mean $\hat{r}$ and $\sigma(r)$ values quoted in Table~\ref{tab:del_performance} are summarized in Fig.~\ref{fig:del_performance}, which also displays results for input maps with $r=0.01$. We do not observe any significant biases compared to the statistical error for a single realization. The small fluctuations in $\hat{r}$ appearing with the spatially-varying foregrounds are related to volume effects in the distributions of the moment expansion parameters, and will be mitigated by using different priors in future work (see Sec.~\ref{section:conclusion}).

The key point of Fig.~\ref{fig:del_performance} is the comparison between the results obtained with and without delensing. For all foreground models and input $r$ values, the average of $\hat{r}$ remains consistent in both cases, thus explicitly verifying that our implementation of delensing does not introduce any additional bias into the parameter estimates. This does not exclude small shifts in $\hat{r}$ occurring for any single realization, as expected given the shrink in the width of the posterior for $r$. Figure~\ref{fig:del_performance} also illustrates the clear error reduction after delensing, with the mean of $\sigma(r)$ decreasing by 37\%, 31\% and 27\% for \texttt{d0s0}, \texttt{d1s1} and \texttt{dmsm}, respectively, in the absence of primordial $B$-modes. In the latter two cases, the four additional sampled parameters result in wider distributions and a more modest delensing-related improvement. Despite the degradation of uncertainties associated with the use of the moment-expansion method, SO's goal precision of $\sigma(r)=0.003$ is achieved even with the most complex of our foreground models. Our results without delensing are broadly consistent with Ref.~\cite{wolz_simons_2023}; the slightly lower $\sigma(r)$ we obtain with the \texttt{d0s0} model can be attributed to the use of optimized pixel weights, while the slightly higher uncertainties and biases with spatially-varying foregrounds are probably related to the wider survey mask. As expected, delensing improvements observed in the present work exceed the $A_{\rm{lens}}=0.5$ forecasts in Ref.~\cite{wolz_simons_2023}, due to our templates achieving a delensing efficiency of $65\%$.

With $r=0.01$ and \texttt{d0s0} foregrounds, the pipeline successfully detects the primordial signal, and the additional cosmic variance due to the presence of tensor modes leads to larger error bars. 
Without delensing, $\sigma(r)$ grows by 16\% compared to the $r=0$ case, reaching a value of $2.2\times 10^{-3}$. This increase is consistent with the results of Ref.~\cite{wolz_simons_2023}. After delensing, we obtain $\sigma(r)=1.5\times 10^{-3}$, corresponding to a 32\% improvement. Note that we do not modify the pixel weights compared to the $r=0$ case (i.e., we do not add the primordial $B$-mode variance to the denominator in Eq.~\ref{eq:weights}), as the value of $r$ will not be known when analyzing real data and the $r=0$ weights will be used by default.

The last column of Table~\ref{tab:del_performance} contains $\sigma(r)$ values obtained after applying the foreground-cleaning pipeline (without delensing) to $r=0$ input maps in which the lensing $B$-mode power was artificially reduced by 65\%. This procedure approximately simulates the effect of delensing at the map level. The good agreement observed between these results and our average uncertainties after cross-spectral delensing is consistent with the theoretical development in Sec.~\ref{section:equivalence} and attests to the robustness of our forecasts.

\section{Conclusion}\label{section:conclusion}

In this work, we implemented delensing in SO's power-spectrum-based foreground-cleaning pipeline and obtained the first performance forecasts for this technique on realistic SO-like maps including Galactic foregrounds and inhomogeneous noise. Our lensing templates, built by optimally coadding quadratic convergence estimators with Gaussian external LSS tracer simulations, allowed to remove 65\% of the lensing $B$-mode power. By treating the template as an additional input channel and including its auto-spectrum as well as all its cross-spectra with the SAT-like maps in the likelihood function, we observed no significant bias and a clear tightening of the posterior distributions for the tensor-to-scalar ratio $r$. An inverse-variance weighting accounting for the lensing $B$-mode power as well as for the noise level in each pixel was applied when computing power spectra, and was shown to outperform the uniform- and hit-count-weighting schemes used previously.

Our parametric foreground model was adjusted to account for the different levels of complexity present in the input simulations. With $r=0$ and uniform dust and synchrotron SEDs, errors on $r$ decreased by 37\% after delensing. When introducing SED spatial variability, the additional parameters required by the moment-expansion method resulted in wider posteriors. Delensing then yielded a 27\% improvement in $\sigma(r)$ and a final value of $2.2\times 10^{-3}$ for our most complex model, well within SO's target sensitivity. Tensor modes at the $r=0.01$ level were successfully detected by the pipeline and led to a 16\% increase in $\sigma(r)$ due to the additional sample variance. We finally validated our results by comparing them to the uncertainties obtained with artificially delensed input maps, thus verifying the equivalence between map-based and cross-spectral delensing.

The most exciting future prospect for this work is of course the application to early SO data as it becomes available within the next year. The main challenge at this stage will be related to timestream filtering of the SAT data, which we will need to account for at the power spectrum level through the use of transfer functions. For the LAT, maximum-likelihood maps will be obtained with a weighting scheme proportional to the Fourier-diagonal inverse detector-detector covariance matrix. This process will also have to be considered in the first step of the pipeline when working with real LAT data. However, as the LAT's first light is not expected to happen before 2025, early demonstrations of delensing will likely be carried out using $B$-mode templates built from external mass tracers and pre-existing ACT or Planck CMB maps. Finally, another aspect we will need to be mindful of when analyzing real data is the choice of cosmology used to generate our theoretical model. An additional free parameter $A_L$ modulating the lensing $B$-mode power spectrum amplitude may be introduced in order to avoid small biases on $r$ related to uncertainties on cosmological parameters. At SO's sensitivity level, the SNR on the CMB lensing power spectrum is expected to be greater than 100~\cite{the_simons_observatory_collaboration_simons_2019} and such uncertainties should therefore be sub-percent. This could be used as a prior on $A_L$ and any inflation of the errors on $r$ should be negligible as a result. 

Other future extensions of this paper will include the use of more realistic LAT-like simulations containing Galactic and extragalactic foregrounds, which have been shown to impact the delensed $B$-mode power spectrum and lead to biases on $r$, both when using internal~\cite{beck_impact_2020, lizancos_impact_2022} and external \cite{lizancos_delensing_2022} tracers of the lensing convergence. By performing map-based foreground cleaning on the LAT-like mock data and quantifying the remaining biases, we will check that these effects can be successfully mitigated as described in Refs~\cite{lizancos_delensing_2022},~\cite{lizancos_impact_2022} and~\cite{beck_impact_2020}. Furthermore, increasing the realism of our analysis will require including more complexity in our external mass tracers, which are currently only generated as Gaussian realizations of theoretical power spectra assuming linear bias.

We may also attempt to combine LAT and SAT $E$-modes when constructing the lensing template, which was shown to improve delensing efficiency by more than 5\% in an idealistic foreground-free situation~\cite{namikawa_simons_2022}. With realistic input maps, this will require a careful investigation of SAT foreground residuals in the cross-spectra between the lensing template and the observed $B$-modes. Regarding our treatment of variable foreground SEDs, one possible improvement would be the implementation of Jeffrey's priors for the moments parameters in order to mitigate volume effects in their posterior distributions. It would also be interesting to implement delensing in the hybrid component separation pipeline described in Ref.~\cite{azzoni_hybrid_2023}. This technique, which starts by removing foregrounds at the map level assuming spatially uniform SEDs and subsequently models the power spectra of the residuals, was found to lead to tighter constraints than the moment expansion used here; upcoming efforts will be dedicated to determining whether this conclusion remains true with delensing.

Finally, it is important to note that all results presented here assume the nominal SO design with three SATs; however, three additional SATs are planned, which will lead to lower noise levels in the sky maps and greater delensing-related improvements on $\sigma(r)$ as indicated by Eq.~\eqref{eq:sigma_r_fisher}. Furthermore, efficient delensing will be essential for upcoming projects such as CMB-S4~\cite{abazajian_cmb-s4_2022}, which requires removing more than 90\% of the lensing $B$-mode power to achieve its target constraints on the tensor-to-scalar ratio. This will only be possible with dedicated high-sensitivity, high-resolution polarization measurements and more optimal, likelihood-based internal-reconstruction techniques~\cite{belkner_cmb-s4_2023}. Delensing will thus be crucial in reaching the full potential of these extremely sensitive future experiments and hopefully unveiling the elusive physics of the early Universe.

\begin{acknowledgments}
The authors would like to thank B.~Yu for providing the simulated maps and power spectra of the external LSS tracers used in this work. We acknowledge the use of the following public software packages: \texttt{CAMB}~\cite{lewis_efficient_2000}, \texttt{healpy}~\cite{zonca_healpy_2019}, \texttt{cmblensplus}~\cite{namikawa_cmblensplus_2021}, \texttt{NaMaster}~\cite{alonso_unified_2019}, \texttt{emcee}~\cite{foreman-mackey_emcee_2012}, \texttt{NumPy}~\cite{harris_array_2020}, \texttt{SciPy}~\cite{virtanen_scipy_2020} and \texttt{matplotlib}~\cite{hunter_matplotlib_2007}. Numerical computations were carried out using resources at NERSC (National Energy Research Scientific Computing Center), a U.S. Department of Energy Office of Science User Facility operated under Contract No. DE-AC02-05CH11231. This work was supported in part by a grant from the Simons Foundation (Award \#457687, B.K.). EH is supported by a Gates Cambridge Scholarship (grant OPP1144 from the Bill \& Melinda Gates Foundation). TN is supported by JSPS KAKENHI Grant No. JP20H05859 and No. JP22K03682. IAC acknowledges support from Fundaci\'on Mauricio y Carlota Botton and the Cambridge International Trust. DA acknowledges support from the Beecroft Trust. CB acknowledges partial support by the Italian Space Agency LiteBIRD Project (ASI Grants No. 2020-9-HH.0 and 2016-24-H.1-2018), as well as the InDark and LiteBIRD Initiative of the National Institute for Nuclear Phyiscs, and the RadioForegroundsPlus Project HORIZON-CL4-2023-SPACE-01, GA 101135036. EC acknowledges support from the European Research Council (ERC) under the European Union’s Horizon 2020 research and innovation programme (Grant agreement No. 849169). AC acknowledges support from the STFC (grant numbers ST/S000623/1 and ST/X006387/1). JE acknowledges the SCIPOL project funded by the European Research Council (ERC) under the European Union’s Horizon 2020 research and innovation program (PI: Josquin Errard, Grant agreement No. 101044073). LP acknowledges the financial support from the COSMOS network (www.cosmosnet.it) through the ASI (Italian Space Agency) Grants 2016-24-H.0 and 2016-24-H.1-2018. BS acknowledges support from
the European Research Council (ERC) under the European Union’s Horizon 2020 research and innovation
programme (Grant agreement No. 851274).
\end{acknowledgments}


\appendix

\section{Hamimeche \& Lewis likelihood}\label{appendix_A}

\begin{figure*}[h!tb]
    \centering
    \begin{minipage}{.48\textwidth}
        \centering
        \includegraphics[width=\linewidth, height=0.25\textheight]{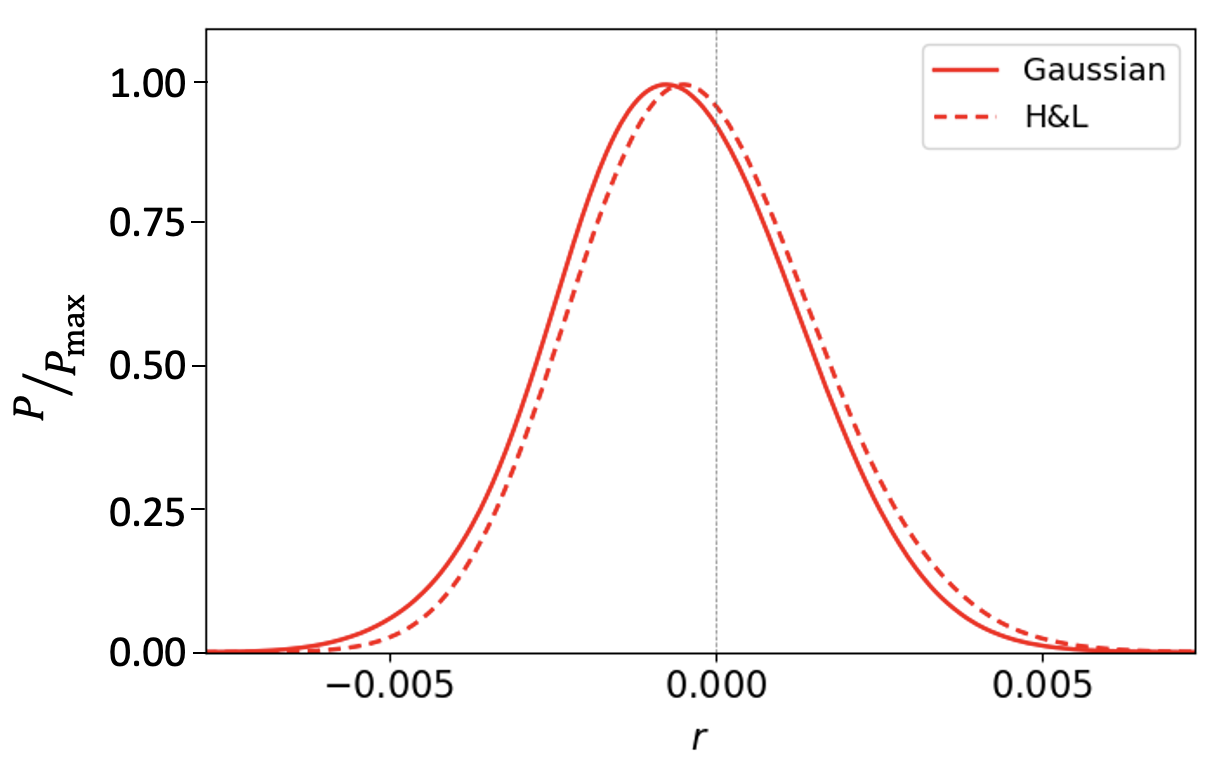}
    \end{minipage}%
    \hfill
    \begin{minipage}{0.48\textwidth}
        \centering
        \includegraphics[width=\linewidth, height=0.25\textheight]{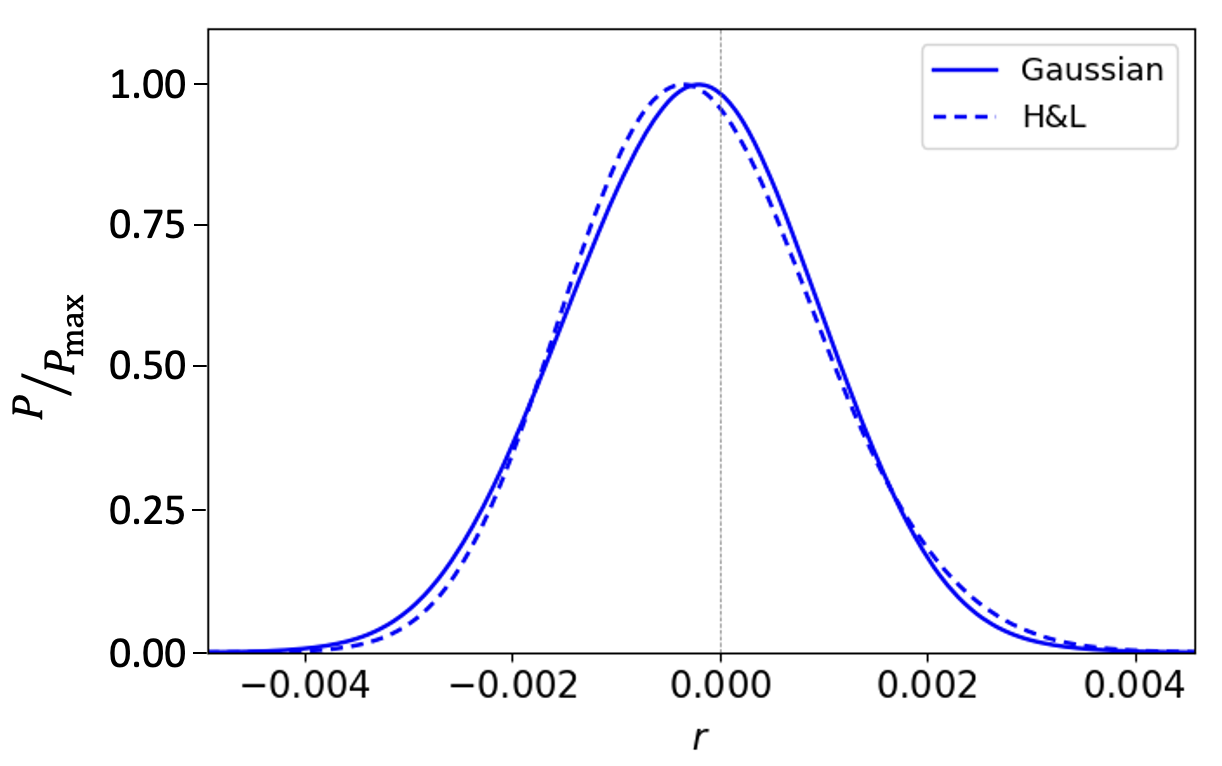}
    \end{minipage}
    \vspace{-5pt}
    \caption{Marginalized posterior distributions for the tensor-to-scalar ratio $r$, obtained by sampling the full Hamimeche \& Lewis likelihood Eq.~\eqref{eq:hnl_final} (dashed lines) and its Gaussian approximation Eq.~\eqref{eq:gauss_likelihood} (solid lines) for a single input realization. The input simulation has $r=0$ and \texttt{d0s0} foregrounds, and cross-spectra are computed using the optimal pixel weights. \textit{Left:} no delensing is applied. \textit{Right:} cross-spectral delensing is performed using the template characterized in Sec.~\ref{section:results_temp}.}
    \label{fig:likelihood_comp}
\end{figure*}

In this appendix we introduce an alternative likelihood to the simple Gaussian approximation used in the main text, following Ref.~\cite{hamimeche_likelihood_2008}. This improved approximation aims to capture the non-Gaussian dependence of the likelihood on the model power spectra. We briefly review the construction of the likelihood and present the results from applying it to one of our simulations with \texttt{d0s0} foregrounds.


The Hamimeche \& Lewis likelihood approximation takes as its starting point the exact likelihood for an ideal, full-sky survey of several correlated Gaussian fields. Considering $n$ fields $\bm{a}_{lm}=(a^{1}_{lm},\ldots,a^{n}_{lm})^T$ (typically seven, with the $B$-modes from each SAT channel and the lensing template), the log-likelihood is given by the matrix generalization of Eq.~\eqref{eq:gaussian}:
\begin{equation}\label{eq:gaussian_matrix}
-\ln{P({\{\bm{a}_{lm}\}|\mathbf{C}_l})}=\frac{2l+1}{2}\left[\textrm{Tr}\left(\hat{\mathbf{C}}_l\mathbf{C}_l^{-1}\right)+\ln{|\mathbf{C}_l|}\right]
\end{equation}
up to a constant,
where $\mathbf{C}_l$ is the model covariance depending on a set of parameters and $\hat{\mathbf{C}}_l=(2l+1)^{-1}\sum_{m=-l}^{l}{\bm{a}_{lm}\bm{a}_{lm}^{\dag}}$. Note that $\bm{a}_{lm}$ contains noise contributions that must be included in the model. Equation~\eqref{eq:gaussian_matrix} is the log-likelihood at a given multipole $l$; the full log-likelihood is obtained by summing over $l$.

In the ideal full-sky likelihood, Eq.~\eqref{eq:gaussian_matrix}, the fields only enter through their power spectra $\hat{\mathbf{C}}_l$. Integrating over the $\bm{a}_{lm}$ at fixed $\hat{\mathbf{C}}_l$, gives the sampling distribution $P(\hat{\mathbf{C}}_l|\mathbf{C}_l)$ for the measured power spectra. The likelihood function for the $\mathbf{C}_l$ given the measured $\hat{\mathbf{C}}_l$ is $\mathcal{L}(\mathbf{C}_l|\hat{\mathbf{C}}_l) = P(\hat{\mathbf{C}}_l|\mathbf{C}_l)$. Normalising the log-likelihood to zero at $\mathbf{C}_l = \hat{\mathbf{C}}_l$, we have~\cite{hamimeche_likelihood_2008}
\begin{equation}\label{eq:wishart}
-\ln{\mathcal{L}({\mathbf{C}_l|\hat{\mathbf{C}}_l})}=\frac{2l+1}{2}\left[\mathrm{Tr}\left(\hat{\mathbf{C}}_l\mathbf{C}_l^{-1}\right)-\ln{\big|\hat{\mathbf{C}}_l\mathbf{C}_l^{-1}\big|}-n\right].
\end{equation}
Introducing a fiducial model with cross-spectra $\mathbf{C}_{fl}$,
Eq.~\eqref{eq:wishart} can be rewritten as
\begin{equation}\label{eq:hnl_matrix}
-\ln{\mathcal{L}({\mathbf{C}_l|\hat{\mathbf{C}}_l})}=\frac{2l+1}{4}\textrm{Tr}\left[\left(\mathbf{C}_{fl}^{-1/2}\mathbf{C}_{gl}\mathbf{C}_{fl}^{-1/2}\right)^2\right]
\end{equation}
with $\mathbf{C}_{gl}=\mathbf{C}_{fl}^{1/2}\mathbf{g}\left(\mathbf{C}_l^{-1/2}\mathbf{\hat{C}}_l\mathbf{C}_l^{-1/2}\right)\mathbf{C}_{fl}^{1/2}$.
Here, the matrix function $\mathbf{g}$ acts on symmetric, positive-definite matrices by application of the function
\begin{equation}
g(x) = \text{sign}(x-1)\sqrt{2(x - \ln x - 1)}   
\end{equation}
to its eigenvalues.
We can now rearrange the upper triangular portion of $\mathbf{C}_{gl}$ into a vector $\bm{X}_{gl}$ in order to express Eq.~\eqref{eq:hnl_matrix} as a quadratic form 
\begin{equation}\label{eq:hnl_1l}
-2\ln{\mathcal{L}({\mathbf{C}_l|\hat{\mathbf{C}}_l})}=\bm{X}_{gl}^T\mathbf{M}_{fl}^{-1}\bm{X}_{gl},
\end{equation}
where $\mathbf{M}_{fl}$ is the covariance of the measured power spectra in the fiducial model.

The exact full-sky likelihood function corresponds to the sum of Eq.~\eqref{eq:hnl_1l} over all considered multipoles. While exact on the full sky, it can also be used as an approximate likelihood for an anisotropic survey substituting for mask-deconvolved measured power spectra in the construction of $\bm{X}_{gl}$ and their covariance for $\mathbf{M}_{fl}$. Mask-induced couplings between multipoles can be accounted for in the covariance, leading to the following final result (known as the Hamimeche \& Lewis likelihood):
\begin{equation}\label{eq:hnl_final}
    -2\ln{\mathcal{L}({\{\mathbf{C}_l\}|\{\hat{\mathbf{C}}_l\}})} \approx \sum_{ll'}{\left[\bm{X}_{g}\right]_l^T\left[\mathbf{M}_{f}^{-1}\right]_{ll'}\left[\bm{X}_{g}\right]_{l'}}.
\end{equation}

In Fig.~\ref{fig:likelihood_comp}, we compare Eq.~\eqref{eq:hnl_final} to its Gaussian approximation Eq.~\eqref{eq:gauss_likelihood} for one realization of our input maps with $r=0$ and \texttt{d0s0} foregrounds. All cross-spectra are computed with the optimal pixel weights in Eq.~\eqref{eq:weights}, and 500 SAT-like simulations with Gaussian foregrounds are used to estimate the covariance matrix. The marginalized posterior distributions obtained for $r$ by sampling these two likelihoods are in good agreement both with and without delensing.

\section{Hybrid weighting method}\label{appendix_B}
As an alternative to the pixel weights discussed in Sec.~\ref{section:weighting}, we investigated a hybrid masking method using both uniform and inverse-noise-variance weighting schemes. To do so, we generated two distinct sets of input maps containing the same realizations of the sky signals (CMB plus Galactic foregrounds) and noise, then applied the uniform mask to one set and the hit-count mask to the other. The same treatment was applied to the lensing $B$-mode template, and the cross-spectra between all pairs of maps (including pairs with different masks) were incorporated in the likelihood. With six frequency bands and the lensing template, the number of coadded spectra therefore increased to 105, compared to the 28 shown in Fig.~\ref{fig:spectra}. 

In Ref.~\cite{efstathiou_myths_2004}, this approach was used to combine pseudo-$C_l$ estimators computed with uniform and inverse-noise-variance weighting into a single hybrid estimator. The advantage of this approach is that it effectively selects different weights on different scales, where isotropic signal or anisotropic noise may dominate. Applying this technique to our analysis would allow us to drop the assumption of white noise, on which the argument in Sec.~\ref{section:weighting} relies, and potentially lead to better results on $r$.

However, issues related to the increased number of cross-spectra appeared in practice: with a covariance matrix obtained from 500 simulations, parameter distributions for an input map containing \texttt{d0s0} foregrounds were unrealistically tight, indicating an inaccurate estimation of the covariance. This problem was shown to stem from the insufficient number of simulations compared to the size of the data vector by performing a simple test with foreground-free input maps (containing only the CMB signal and noise). Restricting the analysis to the three central frequency channels, we obtained a reasonable SD of $\sigma(r)=0.0013$ without delensing (21 cross-spectra) for the hybrid method; using the optimal weights of Sec.~\ref{section:weighting} led to a nearly identical result. With the hybrid method, including all six channels (78 cross-spectra) nearly divided this value by two, even though the low- and ultra-high frequency bands are not expected to contribute significant information in the absence of foregrounds. To obtain accurate error bars with so many cross-spectra, we would need to increase the number of simulations to about 2000 for the covariance estimation. As the results with both weighting methods were equivalent in the simple three-channel test, we concluded that there is little to be gained from the hybrid method for SO over the optimal weights. Given the prohibitive computational cost of accurate covariance estimation for the hybrid method, we decided not to pursue it further. It might, however, be worth revisiting this technique if a future experiment exhibits a clearly non-white noise power spectrum.

\bibliography{apssamp}

\end{document}